\documentclass{JHEP3}

\usepackage{amsmath}
\usepackage{amssymb}
\usepackage{amsfonts}
\usepackage{latexsym}
\usepackage{epsfig}
\usepackage{comment}

\newcommand{\UU}{\mathcal{U}}

\newcommand{\sR}{\mathcal{R}}
\newcommand{\sT}{\mathcal{T}}
\newcommand{\p}{\partial}
\newcommand{\GG}{\mathcal{G}}

\newcommand\hS{{\hat{S}}}
\newcommand{\FF}{\mathcal{F}}

\newcommand{\OO}{\mathcal{O}}
\newcommand{\HH}{\mathcal{H}}

\def\ov{\over}

\def\vev#1{\langle#1\rangle}

\def\th{{\theta}}

\def\Res#1{\mathop{\text{Res}}_{#1}}
\def\sof#1{{\{ {#1} \}}}
\def\eq#1{(\ref{#1})}

\def\vh{{\hat{\varphi}}}

\def\rhat{{\rhat{q}}}
\def\Om{{\Omega}}
\def \th{{\theta}}
\def \Th{{\Theta}}
\def \arccot{{\text{arccot}}}

\def \om {\omega}
\def \ra {\rightarrow}

\def\s{{\sigma}}

\def\LL{{\cal L}}

\def\Ga{{\Gamma}}
\def\ta{{\tilde{a}}}

\def\LL{{\cal L}}

\newcommand{\be}{\begin{equation}}
\newcommand{\ee}{\end{equation}}
\newcommand{\bea}{\begin{eqnarray}}
\newcommand{\eea}{\end{eqnarray}}

\newcommand{\bln}{\begin{align}}
\newcommand{\eln}{\end{align}}
\newcommand{\bst}{\begin{split}}
\newcommand{\est}{\end{split}}
\newcommand{\bi}{\begin{itemize}}
\newcommand{\ei}{\end{itemize}}
\newcommand{\ben}{\begin{enumerate}}
\newcommand{\een}{\end{enumerate}}

\title{Scalar Three-point Functions in a CDL Background}

\author{Daniel S. Park\\
Center for Theoretical Physics\\
Department of Physics\\
Massachusetts Institute of Technology\\
Cambridge, MA 02139, USA\\
\\
\\
{\tt dspark81} {\rm at} {\tt mit.edu}
}

\preprint{MIT-CTP-4316}

\abstract{Motivated by the FRW-CFT proposal by
Freivogel, Sekino, Susskind and Yeh,
we compute the three-point function of
a scalar field in a Coleman-De Luccia instanton
background.
We first compute the three-point function of the
scalar field making only very mild assumptions about the
scalar potential and the instanton background.
We obtain the three-point function for points in the
FRW patch of the CDL instanton and take
two interesting limits;
the limit where the three points are near the boundary of
the hyperbolic slices of the FRW patch,
and the limit where the three points lie on the past lightcone
of the FRW patch.
We expand the past lightcone three-point function
in spherical harmonics.
We show that the near boundary limit expansion of the
three-point function of a massless scalar field
exhibits conformal structure compatible with FRW-CFT
when the FRW patch is flat.
We also compute the three-point function when
the scalar is massive, and explain the obstacles to
generalizing the conjectured field-operator correspondence of
massless fields to massive fields.}

\begin{document}

\section{Introduction} \label{s:intro}

The $AdS/CFT$ correspondence \cite{AdSCFT} has
successfully provided a framework in which to understand
quantum gravity in Anti de Sitter space.
However, an equivalent framework for
gravity in de Sitter space---in which
we live---remains to be understood.
Motivated by the success of $AdS/CFT$,
many holography-inspired ideas
have been put forth on how to
address de Sitter gravity, an incomplete sample
of which has been listed in the bibliography
\cite{dSCFT}-\cite{Symmetree}.
This paper is motivated by one of them,
namely the idea of FRW-CFT
\cite{FSSY, FSSY2, FSSY3, FSSY4, FSSY5}.

\begin{figure}[!ht]
\leavevmode
\begin{center}
\epsfysize=10cm
\epsfbox{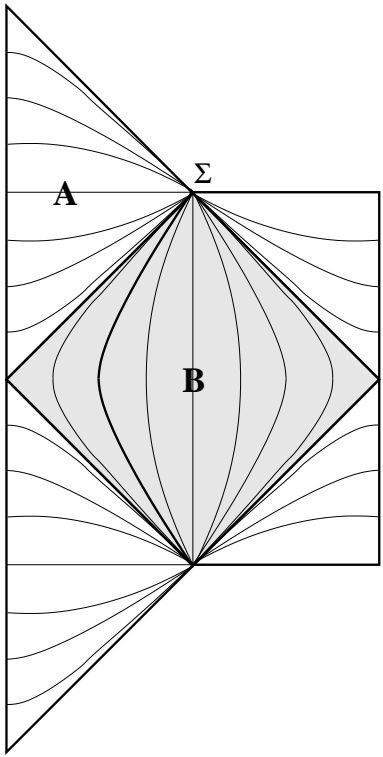}
\end{center}
\caption{\small The Penrose diagram for the Coleman-De Luccia instanton
with an asymptotically flat space inside the bubble
and an asymptotically de Sitter space outside.
In the thin-wall limit, the two regions are divided by a thin domain wall
such as the bold curve in the grey region.
In this case, the space is flat to the left of the wall,
and de Sitter to the right of the wall.
}
\label{BubblePD2}
\end{figure}

The idea of FRW-CFT is that it is natural to
consider quantum gravity in
backgrounds with bubble nucleation,
as described by a Coleman-De Luccia(CDL)
instanton \cite{Coleman:1980aw}.
Suppose we have an asymptotically
flat space inside the bubble
and an asymptotically de Sitter space outside.
The Penrose diagram of this instanton
is shown in figure \ref{BubblePD2}.
If we consider the asymptotically
flat FRW region(region A) of this background in
four dimensions, it has a well defined spatial infinity at $\Sigma$,
which is an $S^2$.

Freivogel, Sekino, Susskind, and Yeh proposed
a holographic correspondence between the
bulk theory in region A and its boundary $\Sigma$ in \cite{FSSY}.
This idea was further elaborated in \cite{FSSY2,FSSY3,FSSY4,FSSY5}.
In these papers, the authors have proposed that in four dimensions,
the holographic dual living at $\Sigma$ corresponding
to the bulk gravity theory is a conformal theory
coupled to a time-like Liouville theory \cite{ST,Schomerus,HMW,Giribet}.
Furthermore, they have identified the
conformal time coordinate with the Liouville field on the boundary.
The full non-perturbative
boundary theory is also conjectured to capture
non-perturbative features of eternal inflation, in particular
the physics of nucleated bubbles
\cite{Satoetal}-\cite{SSS}.
One of the motivations for this conjecture
was the fact that the two-point functions---obtained by
analytic continuation from the Euclidean instanton
to the FRW region of a thin-wall CDL instanton
\cite{GrattonTurok,HertogTurok,HHT}---exhibit
features that suggest the existence of
a holographic CFT.\footnote{An analogous analysis
of two-point functions in general dimensions
was carried out in \cite{Park}.
The late-time behavior of the correlators
in general CDL backgrounds beyond the thin-wall limit
were studied more recently in \cite{DHMST}.}

The two-point function of a single scalar field
that is massless in the FRW region(region A of figure \ref{BubblePD2})
can be decomposed into a tower of two-point functions on
the hyperbolic slices---which are the contours depicted
in figure \ref{BubblePD2}---when the FRW region is flat.
To be more precise,
let us first take the metric in the FRW region to be
\be
ds^2 = e^{2T}(-dT^2 + d\HH^2)
= e^{2T}(-dT^2 + (dR^2+\sinh^2 R d\om^2)) \,,
\ee
where $d\HH^2$ denotes the three-dimensional hyperbolic metric,
and $d\om^2$ denotes the $S^2$ metric.
Then, the two-point functions of massless
scalars can be written in the form
\begin{align}
\begin{split}
G(T,\HH, T',\HH') &= \sum_\Delta e^{-\Delta T} e^{-\Delta T'}G_{\Delta} (\HH,\HH')
+\sum_\Delta e^{(\Delta-2) T} e^{(\Delta-2) T} G_{\Delta} (\HH,\HH') \\
&+  \sum_\Delta e^{-\Delta T} e^{(\Delta-2) T'} G_{\Delta} (\HH,\HH')
+\sum_\Delta e^{(\Delta-2) T} e^{-\Delta T'} G_{\Delta} (\HH,\HH') \,,
\end{split}
\end{align}
in the ``near boundary limit," {\it i.e.,} when $R \ra \infty$.
$G_{\Delta} (\HH,\HH')$ are
two-point functions of dimension $\Delta$
on three-hyperbolic space.
This form made it tempting to conjecture that a massless scalar
in flat FRW space corresponds to
a sum of operators at the boundary of two less dimensions, {\it i.e.,}
\begin{align}
\begin{split}
\phi \rightarrow \sum_\Delta e^{-\Delta T} \OO^1_{\Delta}
+ \sum_\Delta e^{(-2+\Delta) T} \OO^2_{\Delta}
= \sum_{\Delta,\pm} e^{(-1\pm(\Delta-1)) T} \OO^\pm_{\Delta} \,,
\end{split}
\label{fieldop}
\end{align}
with
\be
\vev{\OO_\Delta^\pm (x) \OO_{\Delta'}^\pm (x')}
\propto
{\delta_{\Delta \Delta'} \ov |x-x'|^{2\Delta}}
\ee
in general.

In order to investigate the properties of the conjectured
holographic theory at $\Sigma$, some additional data is needed.
The three-point function is the next object one would naturally
compute to explore the conjectured duality.
This is exactly what we do in this paper.
In this paper, we compute three point functions of scalar fields
in a CDL instanton background and investigate its structure.
As was with the case of the two-point function,
we obtain the three-point function by analytically continuing the
three-point function on the Euclidean CDL instanton.

If there is indeed an FRW-CFT correspondence,
additional information about the ``CFT" should
be encoded in the three-point function.
For example, if a field-operator correspondence such as \eq{fieldop}
were true, one would expect that
the bulk three point functions would have a ``holographic
expansion" of the form
\begin{align}
\begin{split}
<\phi \phi \phi> & =
\sum_\sof{\Delta_i,\s_i}
C^{\s_1,\s_2,\s_3}_{\Delta_1, \Delta_2, \Delta_3}
e^{[-1+\s_1(\Delta_1-1)]T_1} e^{[-1+\s_2(\Delta_2-1)]T_2} e^{[-1+\s_3(\Delta_3-1)]T_3} \\
& \times \UU_{\Delta_1, \Delta_2, \Delta_3} (\HH_1,\HH_2,\HH_3) \,,
\end{split}
\label{holform}
\end{align}
where $\UU_{\Delta_1, \Delta_2, \Delta_3} (\HH_1,\HH_2,\HH_3)$
is a three point function in hyperbolic space with operator dimensions $\Delta_i$.
The sum over $\s_i$ runs over the two signs $(+)$ and $(-)$.
In this case, we can identify
$C^{\pm,\pm,\pm}_{\Delta_1, \Delta_2, \Delta_3}$ as structure coefficients
of the CFT.
One of the main results of the current paper is that
there is indeed an expansion of the form \eq{holform}
of the three-point function
for a scalar field that is massless in the flat FRW region.

We find, however, that the situation is rather different for
scalars with a more general potential in a more general background.
We can write the three-point function of a scalar with
a generic potential in a generic CDL instanton as
\begin{align}
\begin{split}
<\phi \phi \phi> & =
\sum_\sof{\Delta_i}
F_{\Delta_1,\Delta_2,\Delta_3}(T_1,T_2,T_3)
\UU_{\Delta_1, \Delta_2, \Delta_3} (\HH_1,\HH_2,\HH_3) \,.
\end{split}
\label{holform2}
\end{align}
We denote this expansion of the correlator,
the ``holographic expansion."
While the two/three-point correlators of massless scalars
in a flat FRW patch split nicely into four/eight
terms that have definite exponential
scaling with respect to $T$, 
the same is not true in general.
The best we can do for these scalars is to take
$T_i \ra -\infty$ or $T_i \ra \infty$ and examine the behavior of
$F_{\Delta_1,\Delta_2,\Delta_3}(T_1,T_2,T_3)$
at these asymptotic limits.
At early times, we show that
\begin{align}
\begin{split}
<\phi \phi \phi> & \ra
\sum_\sof{\Delta_i}
C_{\Delta_1,\Delta_2,\Delta_3}
e^{-\Delta_1T_1}e^{-\Delta_2T_2}e^{-\Delta_3 T_3}
\UU_{\Delta_1, \Delta_2, \Delta_3} (\HH_1,\HH_2,\HH_3) \,,
\end{split}
\label{holform3}
\end{align}
which is exactly how the three-point function of the
massless scalar, \eq{holform}, behaves at early times.
The late time behavior of the correlator seems
to be non-universal.\footnote{We comment on the holographic expansion
of the massive scalar at late times in the concluding section of this paper.
and leave a careful study of the late-time behavior of correlators
to future work.}
If this is indeed the case,
the behavior of the correlator indicates that
the field-operator correspondence
\eq{fieldop} has to be revised for scalars
that have more general potentials or that
are in more general backgrounds.

Since the procedure of calculating the three-point function
we use is completely general, it can be applied to computing
the three-point correlators of any kind of scalar fluctuation.
In particular, if our universe were inside a nucleated CDL bubble,
this is exactly the calculation one would do to compute the
three-point correlations of a scalar fluctuation observable in the sky.
A particularly interesting limit for these
``observational" purposes can be obtained by taking
\be
T \rightarrow -\infty,\quad R \rightarrow \infty,
\quad T+R = \text{(constant)}\,.
\ee
Light-like trajectories have constant $T+R$, so the
$R$ value of the celestial sphere at some past $T$ is given by
\be
R_\text{past} =(T+R)|_\text{current} -T_\text{past} \,.
\ee
By taking this limit, we obtain three-point functions of points
on the past lightcone of the FRW patch---we have expanded the
three-point functions in spherical harmonics in this limit.
In figure \ref{BubblePD2},
the past lightcone is the boundary of region A and B.
The word ``observational" is put in quotation marks because
although data from the past lightcone is in principle observable,
whether such data can be practically obtained is a completely
different question.

We present the calculation of the three-point function
and its interesting limits in the following way.
As was with the two-point function, the three-point function on the
CDL instanton is obtained by analytically
continuing the Euclidean three-point function.
To make the calculation clear, we carry it out in two steps.
\ben
\item We first calculate the three-point function of a scalar in a
general CDL instanton background.
\begin{itemize}
\item We make only very weak assumptions
about the background and the couplings of the scalar
field to background fields.
In particular, we do not assume the thin-wall limit.
\end{itemize}
\item We examine the two useful expressions of the three-point function.
We write out the holographic expansion,
and also write out the spherical harmonics expansion
on the past lightcone.
\begin{itemize}
\item In general, the holographic expansion \eq{holform2}
cannot be written in the form \eq{holform}. It does, however,
have an exponential $T$ scaling \eq{holform3} in the early-time limit,
{\it i.e.,} when $T \rightarrow -\infty$.
\end{itemize}
\een

The structure of the three-point function is determined entirely
by data that can be extracted from the radial profile of the CDL instanton.
We identify the corresponding data for two examples.
\ben
\item A scalar field that is massless in the flat region.
\begin{itemize}
\item We show that the the holographic expansion
can be written in the form \eq{holform} and compute
its structure coefficients.
\end{itemize}
\item A massive scalar field.
\begin{itemize}
\item We compute the data relevant to the three-point function,
and obtain the early-time holographic expansion whose
terms have exponential scaling with respect to $T$.
\end{itemize}
\een
In both examples we set the background to be a thin-wall CDL instanton
considered in \cite{FSSY}, where space is flat on one side
and de Sitter on the other.
We compute the three-point function in the flat FRW region.

The organization of this paper is as follows.
First, we explain the general setup in which we work in
section \ref{s:objective}.
In particular, we review some relevant facts about
the CDL instanton and conformal coordinates
that we refer to throughout the paper.
We calculate the Euclidean three-point function in section \ref{s:euc}.
We analytically continue the three-point function to Lorentzian signature
in section \ref{s:ancon}.
We also take the two useful limits of the analytically continued
three-point function in this section;
the near boundary limit, in which we write the
holographic series expansion of the three-point function,
and the past lightcone limit, in which we expand it in terms of spherical harmonics.

We work out examples in the next two sections.
As noted above, we assume a thin-wall CDL instanton
for explicit calculation for both examples.
In section \ref{s:m0thinwall} we examine the three-point function
in the case that the scalar field is massless in the flat region
after reviewing the thin-wall CDL instanton.
In section \ref{s:m1thinwall} we examine
the case when the scalar field is massive.
Finally in section \ref{s:conclusion} we summarize the
results and discuss its implications in the context of FRW-CFT.

We recommend that the busy reader focus on
sections \ref{s:objective} and \ref{s:conclusion}.
We have put effort into structuring these two sections
so that they present a self-contained summary of the
setup and results of this paper.

\section{The Setup} \label{s:objective}

We wish to compute the three-point function
of a massive scalar around a Coleman-De Luccia instanton.
The Euclidean Lagrangian is given by
\be
\LL = \int d^4x \sqrt{g} (R + {1 \ov 2}(\p \phi)^2 + V(\phi)
+{1 \ov 2}(\p \vh)^2 + \mathcal{V}(\vh,\phi))\,.
\ee
where $\phi$ is the tunneling scalar and
$\vh$ is the scalar whose three-point function we
wish to compute.
We assume $V(\phi)$ have two local minima at $\phi = \phi_\pm$
with values $V(\phi_+)>0$ and $0 = V(\phi_-) < V(\phi_+)$.
We assume that the global minimum of $\mathcal{V}(\vh,\phi)$ is at
$\vh=0$ independent of $\phi$, which implies that
\be
{\delta \mathcal{V} \ov \delta \vh} |_{\vh=0}=
{\delta \mathcal{V} \ov \delta \phi} |_{\vh=0} =0 \,.
\ee
These equalities imply that
there is no mixing between $\phi$ and $\vh$, {\it i.e.,}
\be
{\delta^2 \mathcal{V} \ov \delta \phi \delta \vh} |_{\vh=0} =0 \,.
\ee

The Euclidean CDL instanton solution is given by
\cite{Coleman:1980aw}
\begin{align}
ds^2 &= dr^2 + f(r)^2 d\Om^2 \\
\phi &= \phi_0 (r) 
\end{align}
where we have used $d\Om^2$ to denote the metric on $S^3$.
We use $\Om$ to denote $S^3$ coordinates throughout this paper.
$f(r)$ and $\phi_0 (r)$ must satisfy
\begin{align}
{\phi_0} '' + 3 { f' \ov f} \phi_0' &= {dV \ov d\phi}|_{\phi=\phi_0} \\
{f'}^2 &= 1 + {f^2 \ov 6} \left( {1 \ov 2} {\phi_0' }^2 - V(\phi) \right)
\end{align}
where the primes denote differentiation with respect to $r$.
We demand that $\phi_0$ interpolates from $\phi_+$ to
$\phi_t$(which is between $\phi_+$ and $\phi_-$)
from $r_0$ to $0$ such that
\be
f'(0) =1,\quad f'(r_0) =-1,\quad \phi_0' (0) = \phi_0' (r_0) = 0\,.
\ee

It is useful to use the conformal coordinate $X$, {\it i.e.,}
\be
dX \equiv dr/f(r)
\ee
and to define
\be
a(X) \equiv f(r) \,.
\ee

\begin{figure}[!t]
\centering\includegraphics[width=6cm]{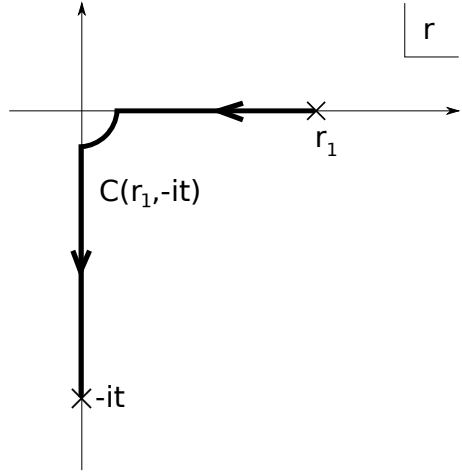}
\caption{\small Contour of integration for conformal coordinates.}
\label{f:radialcoord}
\end{figure}

We would like to analytically continue the coordinates into
the FRW region inside the bubble
\cite{GrattonTurok}.\footnote{We have found it convenient to use
conventions that differ by a sign from \cite{GrattonTurok}.
We briefly explain this choice in section \ref{ss:thinwall}.}
We choose to analytically continue by
\be
r \rightarrow -it, \qquad \theta \rightarrow iR\,.
\ee
Let us analytically continue the conformal radial
coordinate accordingly.
The conformal coordinate $X$ is defined as
\be
X(r) = \int_{r_1}^{r} {dr' \ov f(r')} + X(r_1)
\ee
for some $r_1$.
Since $f(r) \rightarrow r$ as $r \rightarrow 0$,
$X \rightarrow \ln r$ as $r \rightarrow 0$.
The conformal radial coordinate $X$ in the FRW region can be defined as
\be
X(-it) = \int_{C(r_1,-it)} {dr' \ov f(r')} + X(r_1)
\ee
where $C(r_1,-it)$ is a contour that begins at $r_1$ and ends at $-it$
as depicted in figure \ref{f:radialcoord}.

Note that as $t \rightarrow 0$, since $f(r) \rightarrow r$ as $r \rightarrow 0$,
when $t \rightarrow 0$,
\be
X(t) \sim \ln (-it) = \ln t - i\pi/2 \,.
\ee
Now $f(-it)=-i \tilde{f}(t)$ for some real function $\tilde{f}$.
Hence we find that
\be
X(-it) = T(t) -i\pi/2
\ee
for some real function $T$. We define this function to be the
conformal time coordinate.
One finds that for
\be
\tilde{a}(T(t)) \equiv \tilde{f}(t)= if(-it)= i a(X(-it)) =ia(T-i\pi/2)\,,
\ee
the metric becomes
\be
ds^2 = \ta(T)^2 (-dT^2 + d \HH^2)
\ee
where $d\HH^2$ is the metric on the hyperbolic space $\HH^3$.

Note that in the $X \rightarrow -\infty$ limit,
$a(X) \rightarrow L e^X$ for a length scale $L$.
In the $T \rightarrow -\infty$ limit
$\ta(T) \rightarrow L e^T$ for the same $L$.
The analytic continuation is done by joining the Euclidean instanton
and the Lorentzian background at the ``hip," by gluing
$T \ra -\infty$ and $X \ra -\infty$.
We explicitly work this ``patching" out for the thin-wall case
at the beginning of section \ref{s:m0thinwall}.

Now let us expand the fluctuation of $\vh$ around this solution.
Defining
\be
\vh(X,\Om) = a(X)^{-1} \varphi (X,\Om)
\ee
we obtain
\be
\LL =  \int dX d\Om_3 \sqrt{g_{S^3}}
\left(
{1 \ov 2} \varphi (-\p_X^2 + U(X) - \Box) \varphi + {1 \ov 6} W(X) \varphi^3
\right) + \OO(\varphi^4)
\ee
where $g_{S^3}$ refers to the $S^3$ metric
and $\Box$ is the Laplacian on $S^3$.
$U$ and $W$ are given by
\begin{align}
U(X) &= {a'' (X) \ov a(X) }
+a^2 (X) {\delta^2 \mathcal{V} \ov \delta \vh^2}|_{\phi=\phi_0, \vh=0}\\
W(X) &=
a(X) {\delta^3 \mathcal{V} \ov \delta \vh^3}|_{\phi=\phi_0, \vh=0}
\end{align}
We refer to $U(X)$ as the ``radial potential" throughout
this paper.

Our aim is to obtain an expression for
\be
\langle
\varphi (X_1, \Om_1 )
\varphi (X_2, \Om_2 )
\varphi (X_3, \Om_3 )
\rangle
\ee
and its analytic continuation to Lorentzian signature.
The three-point function for $\vh$ can be recovered
by multiplying factors of $a(X_i)^{-1}$ to the correlator of $\varphi$.

Later on, we consider two examples of potentials $\mathcal{V}$.
We first consider a potential for which $\vh$ is massless on one side of
the wall at $X=X_0$ in the thin-wall limit of the instanton.
For example, the potential
\be
\mathcal{V} = {1 \ov 2} {m^2 \ov (\phi_+-\phi_-)^2}
(\phi-\phi_-)^2 \vh^2
+ {1 \ov 6} \lambda {(\phi-\phi_-) \ov (\phi_+-\phi_-)}\vh^3
+ \OO(\vh^4) \,.
\label{pot1}
\ee
serves the purpose as $\phi=\phi_-$ on one side of the thin
wall and $\phi=\phi_+$ on the other.
\begin{align}
U(X) &= \begin{cases}
{a''(X) \ov a(X)} & (X <X_0) \\
{a''(X) \ov a(X)} +m^2 a(X)^2 & (X >X_0)
\end{cases} \\
W(X) &= \begin{cases}
0 & (X <X_0) \\
\lambda a(X) & (X >X_0)
\end{cases}
\end{align}
We refer to a scalar with this potential
as a ``massless scalar" throughout
the paper, for lack of a better term.

We also consider the case when $\mathcal{V}$ is independent of
$\phi$, {\it i.e.,}
\be
\mathcal{V} = {1 \ov 2} m^2 \vh^2
+ {1 \ov 6} \lambda \vh^3
+ \OO(\varphi^4) \,.
\label{pot2}
\ee
The scalar must be massive in order for
$\vh=0$ to be a stable point.
Then
\begin{align}
U(X) &= {a''(X) \ov a(X)} +m^2 a(X)^2 \\
W(X) &= \lambda a(X)
\end{align}

\section{Correlators in Euclidean Signature} \label{s:euc}

In this section, we compute the two-point and three-point
correlators on the Euclidean CDL instanton.
We review the computation of the two-point function in section
\ref{ss:euc2} for a general CDL instanton, assuming
only mild conditions on the properties of the radial potential
\be
\label{U}
U(X) = {a'' (X) \ov a(X) }
+a^2 (X) {\delta^2 \mathcal{V} \ov \delta \vh^2}|_{\phi=\phi_0, \vh=0}\,.
\ee
Using the two-point function, we write an expression for
the three-point function in section \ref{ss:euc3}.

\subsection{The Two-Point Function} \label{ss:euc2}

We write out the two-point function in a form convenient for
our purposes in this section.
A more detailed account of the calculation can
be found in \cite{FSSY,GrattonTurok,HertogTurok,HHT,Park}.
Many of the results on one-dimensional scattering used in this
section can be found in \cite{FSSY,Barton}.

Let us consider the Schr\"odinger equation,
\be
(-\p_X^2 + U(X)) \Psi_E (X) = E \Psi_E (X)\,,
\ee
where $U$ is the radial potential, \eq{U}.
By properties of $a(X)$, it is easy to verify that
\be
U(X) \rightarrow 1 \quad \text{for } X\rightarrow \pm \infty\,.
\ee
So there exist a continuum of states labelled by real number $k$
that satisfy
\be
(-\p_X^2 + U(X)) \Psi_k (X) = (k^2+1) \Psi_k (X)\,.
\ee

There are two different bases that we can organize such solutions into.
We define $\Psi_k$ to be the solutions that behave asymptotically as
\begin{align}
\Psi_k &\rightarrow \begin{cases}
e^{ikX} + \sR(k) e^{-ikX}   & (X \rightarrow -\infty) \\
\sT(k) e^{ikX}  & (X \rightarrow \infty)
\end{cases}
\end{align}
Also, we define $\Phi_k$ to be the solutions that behave asymptotically as
\begin{align}
\Phi_k &\rightarrow \begin{cases}
\sT(k) e^{-ikX}   & (X \rightarrow -\infty) \\
e^{-ikX} - {\sR(-k) \sT(k) \ov \sT(-k)} e^{ikX}   & (X \rightarrow \infty)
\end{cases}
\label{Philim}
\end{align}

We note that for real $k$
\be
\Psi_{-k} = \Psi_{k}^*,\quad\Phi_{-k} = \Phi_{k}^*\,,
\ee
and hence that
\be
\sR(k)^*=\sR(-k),\quad \sT(k)^*=\sT(-k)\,.
\ee
The unitarity relation $|\sR|^2 + |\sT|^2 =1$ becomes
\be
\sR(k)\sR(-k) + \sT(k)\sT(-k) =1\,.
\ee
Also the two bases are related by
\begin{align}
\Psi_k &= {1\ov \sT(-k)} \Phi_{-k} + {\sR(k) \ov \sT(k)} \Phi_{k} \label{psiphirel1}\\
\Phi_k &= {1\ov \sT(-k)} \Psi_{-k} - {\sR(-k) \ov \sT(-k)} \Psi_{k}\,. \label{psiphirel2}
\end{align}

There also can be bound states of this potential.
We already know that
\be
U(X) \rightarrow 1 \quad \text{for } X\rightarrow \pm \infty\,.
\ee
In addition to this, we assume that the following holds:
\ben
\item The poles of $\Phi_k$, $\Psi_k$
and $\sT(k)$ with respect to $k$
in the upper-half of the complex $k$ plane coincide
and are simple.
\item The number of such poles are finite.
\item All such poles $iz$ lie on the
imaginary axis and correspond to unique bound states
of energy $(1-z^2)$.
\item $\Phi_k / \sT(k)$ does not have a pole in the
upper-half of the complex $k$ plane.
\item $U(X)$ approaches $1$ as $X \ra -\infty$ ``rapidly."
\een
We refer to these conditions as ``regularity conditions"
throughout this paper.
We have defined the meaning of ``rapid" in appendix
\ref{ap:complete}.
For our purposes, it is enough to note that exponential
tails---($U(X) \sim 1 + Ae^{NX}$)---are ``rapid" enough.

The second condition means that for each pole
$iz$ of $\sT$ in the upper-half of the
complex $k$ plane, there exists a bound state
\be
u_{iz} (X) \propto \Res{k=iz} \Psi_k \propto \Res{k=iz} \Phi_k 
\ee 
such that
\be
(-\p_X^2 + U(X)) u_{iz} (X) = (-z^2+1) u_{iz} (X)\,.
\ee
All bound states are non-degenerate and hence their wavefunctions
are real up to overall phase. We can fix the phase to be $0$.

These assumptions all hold when the potential $U(X)$
becomes constant for $X<X_B$ for some $X_B$.
In that case, all the poles of the reflection coefficient $\sR$
and transmission coefficient $\sT$ coincide in the upper-half of
the $k$ plane.
Also, all these poles lie on the imaginary axis and correspond
to unique bound states.
We have slightly generalized these restrictions in our case.
Rather than restricting the analytic structure of the scattering
coefficients, we have restricted the analytic structure of
the eigenfunctions themselves.
We believe that these assumptions are not very strong,
as we expect $\Phi, \Psi$ and $\sT$ of a 
``generic" $U(X)$ to obey this property.
We note that poles of $\Phi_k, \Psi_k$ and $\sT_k$
in the lower-half plane are relatively uncontrollable
in contrast to the poles in the upper-half plane.

\begin{figure}[!t]
\centering\includegraphics[width=6cm]{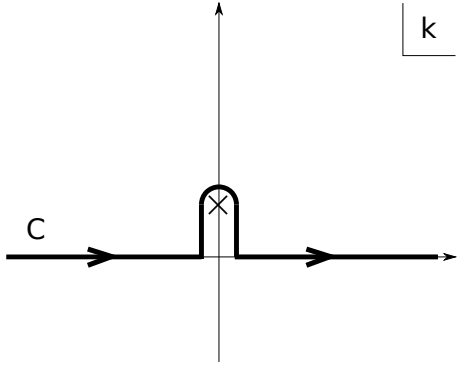}
\caption{\small The contour $C$. $C$ runs along the real axis with a 
jump over the upmost pole of $\sT$ marked by a cross.}
\label{f:contourp}
\end{figure}

Defining $C$ to be a contour in the complex $k$ plane
that runs along the real axis with a jump over all the poles
of $\sT$ in the upper-half plane,
the two-point function
is given by \cite{GrattonTurok,HertogTurok,HHT}
\begin{align}
\begin{split}
G((X_1, \Om_1 ), (X_2, \Om_2 )) &\equiv
\langle
\varphi (X_1, \Om_1 )
\varphi (X_2, \Om_2 )
\rangle \\
&= \int_C {dk \ov 2\pi} {\Phi_{k}(X) \Psi_{k} (X') \ov \sT(k)} G_k (\Om,\Om') \,.
\label{2pt}
\end{split}
\end{align}
The contour $C$ is depicted in figure \ref{f:contourp}.
$G_k$ is the propagator for a scalar
\be
(\Box-(k^2+1))G_{k}(\Om , \Om') = {1 \ov \sqrt{g_{S^3}}} \delta (\Om, \Om') \,.
\ee
Written out explicitly, it is \cite{FSSY}
\be
G_{k_i}(\Om , \Om') = {\sinh k_i (\pi -\Th) \ov \sinh k_i \pi \sin \Th } \,.
\ee
$\Th(\Om,\Om')$
is the angular distance from $\Om$ to $\Om'$.

To show \eq{2pt}, we need to show that
\be
\int_C {dk \ov 2\pi} {\Phi_{k}(X) \Psi_{k} (X') \ov \sT(k)}  = \delta (X-X') \,. \label{delta}
\ee
If this were true, then
\begin{align}
\begin{split}
&(-\p_X^2 + U(X) - \Box)  \int_C {dk \ov 2\pi} {\Phi_{-k}(X) \Psi_{k} (X') \ov \sT(k)} G_k (\Om,\Om') \\
&= \int_C {dk \ov 2\pi} (-\p_X^2 + U(X)){\Phi_{-k}(X) \Psi_{k} (X') \ov \sT(k)} G_k (\Om,\Om') \\
& \qquad\qquad\qquad\qquad\qquad
- \int_C {dk \ov 2\pi} {\Phi_{-k}(X) \Psi_{k} (X') \ov \sT(k)} \Box G_k (\Om,\Om') \\
&= \int_C {dk \ov 2\pi} {\Phi_{-k}(X) \Psi_{k} (X') \ov \sT(k)} (k^2+1-\Box) G_k (\Om,\Om')  \\
&= \int_C {dk \ov 2\pi} {\Phi_{-k}(X) \Psi_{k} (X') \ov \sT(k)} {1 \ov \sqrt{g_{S^3}}} \delta (\Om, \Om') \\
&=  {1 \ov \sqrt{g_{S^3}}} \delta(X-X') \delta (\Om, \Om') 
\end{split}
\end{align}
as desired. The completeness relation \eq{delta} is proven in
appendix \ref{ap:complete}.

To summarize, the two-point function on the Euclidean
CDL instanton can be written in the form
\[
\fbox{
\addtolength{\linewidth}{-2\fboxsep}%
\addtolength{\linewidth}{-2\fboxrule}%
\begin{minipage}{\linewidth}
\begin{align}
\begin{split}
\langle
\varphi (X_1, \Om_1 )
\varphi (X_2, \Om_2 )
\rangle
= \int_C {dk \ov 2\pi} {\Phi_{k}(X) \Psi_{k} (X') \ov \sT(k)} G_k (\Om,\Om')
\label{2pt}
\end{split}
\end{align}
\smallskip
\end{minipage}\nonumber
}
\]
when the potential $U(X)$ satisfies the regularity conditions.
The contour of integration is defined to be a contour that 
runs along the real axis with a jump over the poles of $\sT$;
it is depicted in figure \ref{f:contourp}.
We note once more that $U(X) \sim Ae^{NX}+1$
approaches $1$ in the limit $X \ra -\infty$ rapidly enough
to be regular.

\subsection{The Three-Point Function} \label{ss:euc3}

We find an expression for the tree-level three-point function
\be
T(\sof{X_i},\sof{\Om_i}) \equiv
\langle
\varphi (X_1, \Om_1 )
\varphi (X_2, \Om_2 )
\varphi (X_3, \Om_3 )
\rangle
\ee
in this section.
For notational simplicity, we use $\{ x_i \}$ to denote the triplet
$(x_1, x_2, x_3)$ for any argument $x_i$ throughout the paper.
Recall that the relevant part of the Lagrangian
for computing the three-point function is
\be
\LL =  \int dX d\Om_3 \sqrt{g_{S^3}}
\left(
{1 \ov 2} \varphi (-\p_X^2 + U(X) - \Box) \varphi + {1 \ov 6} W(X) \varphi^3
\right) + \OO(\varphi^4)
\ee
and therefore
\begin{align}
\begin{split}
&T(\sof{X_i},\sof{\Om_i})  \\
&= \int dX d\Om W(X) G((X_1, \Om_1 ),(X,\Om))
G((X_2, \Om_2 ),(X,\Om))
G((X_3, \Om_3 ),(X,\Om))
\end{split}
\end{align}
at tree-level.

Using \eq{2pt}, we find that
the full three-point function of $\varphi$
on the Euclidean instanton is given by
\begin{align}
\begin{split}
& T(\sof{X_i},\sof{\Om_i}) \\
&=\int dX W(X)
\int d \Om \prod_i
\left( \int_C {dk_i \ov 2\pi} {\Phi_{k_i} (X_i) \ov \sT(k_i)} \Psi_{k_i}(X) G_{k_i} (\Om, \Om_i)\right) \\
&=\left( \prod_i  \int_C {dk_i \ov 2\pi} {\Phi_{k_i} (X_i) \ov \sT(k_i)}  \right)
\left( \int dX W(X) \prod_i \Psi_{k_i}(X) \right)
U_\sof{k_i} (\sof{\Om_i}) \\
&=\left( \prod_i  \int_C {dk_i \ov 2\pi} {\Phi_{k_i} (X_i) \ov \sT(k_i)}  \right)
S(\sof{k_i}) U_\sof{k_i} (\sof{\Om_i}) \,.
\end{split}
\label{full0}
\end{align}
Here we have defined
\be
S(\sof{k_i}) =  \int_{-\infty}^\infty dX W(X) \Psi_{k_1}(X)\Psi_{k_2}(X)\Psi_{k_3}(X) \,,
\ee
and
\be
U_{k_1,k_2,k_3} (\Om_1, \Om_2, \Om_3)
\equiv \int d \Om_0 G_{k_1}(\Om_1 , \Om_0)
G_{k_2}(\Om_2 , \Om_0) G_{k_3}(\Om_3 , \Om_0)\,.
\ee
As will be seen throughout this paper, $S(\sof{k_i})$ is the crucial data
that determines the three-point function. We call $S$ the ``wavefunction overlap."

\section{Analytic Continuation of the Three-Point Function} \label{s:ancon}

We analytically continue the Euclidean CDL three-point function
to Lorentzian signature in this section and take various useful limits.
As a first step, we analytically continue the three-point function
on $S^3$ to $\HH^3$ in section  \ref{ss:ancons3}.\footnote{This
analytic continuation was the key step that made this paper possible.
The results of section \ref{ss:ancons3} were obtained jointly with Yasuhiro Sekino,
and I am indebted to him for providing crucial insight to this calculation.}
Next, we use this result to write the holographic expansion of the
CDL three-point function in section \ref{ss:holexp}.

Lastly, we write the three-point function for points
lying on the past lightcone of the FRW region in section \ref{ss:harm}.
Recall that the past lightcone is at
\be
T \rightarrow -\infty,\quad R \rightarrow \infty, \quad T+R = \text{(constant)}\,
\ee
as seen in the introduction.
We expand the correlators in terms of
$S^2$ harmonics.

\subsection{Analytic Continuation of the Three-point Function on $S^3$} \label{ss:ancons3}

We first must understand how to analytically
continue the three-point function on $S^3$
of three scalars $\phi_1, \phi_2, \phi_3$ that satisfy
\be
(\Box - (k_i^2+1)) \phi_i =0
\ee
with the interaction term
\be
\phi_1 \phi_2 \phi_3 \,.
\ee

We use the coordinates $(\theta,\phi,\varphi)$
on $S^3$ where $\th$ and $\phi$ vary from $0$ to $\pi$
and the range of $\varphi$ is given by $[0,2\pi)$.
The metric on $S^3$ is given by
\be
ds^2 = d\theta^2 + \sin^2 \theta d\om^2
\equiv d\theta^2 + \sin^2 \theta (d\phi^2 + \sin^2 \phi d \varphi^2 )  \,.
\ee
We use $\Om$ as shorthand for the $S^3$
coordinates $(\theta,\phi,\varphi)$ and $\om$
as shorthand for the $S^2$ coordinates $(\phi,\varphi)$.
The propagator for each scalar satisfying
\be
(\Box-(k_i^2+1))G_{k_i}(\Om , \Om') = {1 \ov \sqrt{g_{S^3}}} \delta (\Om, \Om')
\ee
is given by \cite{FSSY}
\be
G_{k_i}(\Om , \Om') = {\sinh k_i (\pi -\Th) \ov \sinh k_i \pi \sin \Th } \,.
\ee
$\Th(\Om,\Om')$
is the angular distance from $\Om$ to $\Om'$;
\begin{align}
\cos \Th(\Om,\Om')
= \cos \th \cos \th' + \sin \th \sin \th' \cos \alpha(\om,\om') \,.
\end{align}
Here $\alpha$ is the angular distance between the $S^2$
coordinates;
\be
\cos \alpha(\om,\om') =
\cos \phi \cos \phi' + \sin \phi \sin \phi' \cos (\varphi -\varphi') \,.
\ee

The three-point function is given by
\be
U_{k_1,k_2,k_3} (\Om_1, \Om_2, \Om_3)
= \int d \Om_0 G_{k_1}(\Om_1 , \Om_0)
G_{k_2}(\Om_2 , \Om_0) G_{k_3}(\Om_3 , \Om_0) \,,
\ee
which we have encountered at the end of section \ref{ss:euc3}.
Using the shorthand notation
\be
\Th_{ij} = \Th (\Om_i , \Om_j) \,,
\ee
this can be written as
\begin{align}
U_{k_1,k_2,k_3} (\Om_1, \Om_2, \Om_3) 
= \int_0^\pi d \theta_0 \sin^2 \th_0 \int d \om_0
\prod_i {\sinh k_1 (\pi -\Th_{i0}) \ov  \sinh (k_i \pi )\sin \Th_{i0} }
\,.
\end{align}
As before, using the notation $\{ x_i \}$
to denote the triplet $(x_1, x_2, x_3)$ for
any argument $x_i$, $U$ can be
conveniently written as
\be
U_{\sof{ k_i}} (\sof{\Om_i})
= U_{k_1,k_2,k_3} (\Om_1, \Om_2, \Om_3) 
 \,.
\ee

Let us analytically continue $\th$ to $iR$.
Then
\be
-ds^2 =dR^2 + \sinh^2 R d\om^2
\ee
is precisely the metric on hyperbolic space $\HH^3$.
Interpreting the coordinates $(R,\phi,\varphi)$ as coordinates
on $\HH^3$ we may analytically continue the arguments
of $T$ to obtain a function on $\HH^3$;
\begin{align}
u_\sof{k_i} (\sof{(R_i,\phi_i,\varphi_i)})
\equiv U_\sof{k_i} (\sof{(iR_i,\phi_i,\varphi_i)})
\,.
\end{align}
We use $\HH$ to denote the coordinates $(R,\phi,\varphi)$.

\begin{figure}[!t]
\centering \includegraphics[width=6cm]{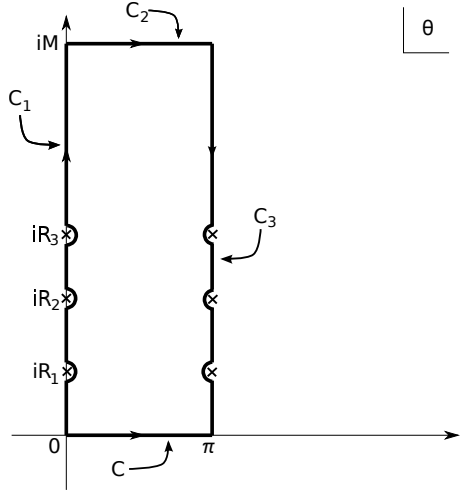}
\caption{\small Contour of integration for $\theta_0$.}\label{f:contourR}
\end{figure}

Now we deform the initial contour of integration $C$
for $\th_0$ on the complex plane.
The only potential poles are at $\th = iR_i + n \pi$
for integer $n$ so we deform the contour of integration as in figure \ref{f:contourR}
and take $M$ to infinity.
The new contour of integration can be broken in to 3 pieces $C_1$,
$C_2$ and $C_3$, {\it i.e.,}
\be
u_\sof{k_i} (\sof{\HH_i}) = u^1_\sof{k_i} (\sof{\HH_i}) 
+u^2_\sof{k_i} (\sof{\HH_i})  + u^3_\sof{k_i} (\sof{\HH_i}) \,,
\ee
where
\be
u^I_\sof{k_i} (\sof{\HH_i}) 
= \int_{C_I} d \theta_0 \sin^2 \th_0 \int d \om_0
\prod_i {\sinh k_1 (\pi -\Th_{i0}) \ov  \sinh (k_i \pi )\sin \Th_{i0} } \,.
\ee

As we take $M$ to infinity the integral over $C_2$ vanishes.
This is because when $\Om_0 = (iM+\th_0,\phi_0,\varphi_0)$
\begin{align}
\begin{split}
&\cos \Th_{i0} \\
&={1 \ov 2} e^M ( \cos \th_0(\cosh R_i + \sinh R_i \cos \alpha_{0i})
+i \sin \th_0(\cosh R_i - \sinh R_i \cos \alpha_{0i})) + \OO(e^{-M}) \\
&={1 \ov 2} e^{(M+\OO(1))} (\cos \th + i \sin \th)+ \OO(e^{-M}) \,,
\end{split}
\end{align}
where
\be
\cot \th = \left( {\coth R_i + \cos \alpha_{0i} \ov \coth R_i - \cos \alpha_{0i} } \right) \cot \th_0 \,.
\ee
Hence
\be
\arccot (e^{2R_i} \cot \th_0) \leq  \th  \leq \arccot (e^{-2R_i} \cot \th_0) \,.
\ee
It is clear that $\th$ slides from $0$ to $\pi$
as $\th_0$ runs from $0$ to $\pi$.
Since
\be
\cos (-iM + \theta ) ={1 \ov 2} e^{(M+\OO(1))} (\cos \th + i \sin \th)+ \OO(e^{-M})
\ee
for large $M$, it is clear that up to corrections small in the limit of large $M$,
\be
\Th_{i0} = -iM + \theta
\ee
where $\theta$ ranges from $0$ to $\pi$.
Therefore
\be
|\sin \Th_{i0} | \simeq {1 \ov 2} e^M 
\ee
for large $M$ and
\be
\left| \sinh k_i (\pi- \Th_{i0}) \right|
\ee
is bounded above for given $\HH_1, \HH_2$ and $\HH_3$
for $\th_0$ on $C_2$ since the real part of $\Th_{i0}$ has finite range.
Therefore
\begin{align}
\left|\int_{C_2} d \theta_0 \sin^2 \th_0 \int d \om_0
\prod_i {\sinh k_1 (\pi -\Th_{i0}) \ov  \sinh (k_i \pi )\sin \Th_{i0} } \right|
\sim \OO (e^{-M})
\end{align}
and hence $u^2_\sof{k_i} (\sof{\HH_i})$ vanishes as $M \rightarrow \infty$.

Now let us carry out the integral along the contours $C_1$ and $C_3$.
\begin{align}
\begin{split}
\cos \Th ((iR,\om),(iR',\om'))
&= \cosh R \cosh R' - \sinh R \sinh R' \cos \alpha(\om,\om') \\
&= \cosh \ell(\HH, \HH')
\end{split}
\end{align}
where $\ell(\HH, \HH')$ is the angular distance between two
points in hyperbolic space. Therefore
\begin{align}
\begin{split}
G_{k_j} ((iR_0,\om_0),(iR_j,\om_j))
&={\sinh k_j (\pi-i \ell_{j0}) \ov \sinh (k_j \pi) \sin i \ell_{j0}}\\
&={e^{ k_j \pi}e^{-i k_j \ell_{j0}}-e^{- k_j \pi}e^{ik_j \ell_{j0}} \ov 2i \sinh (k_j \pi) \sinh \ell_{j0}} \,,
\end{split}
\end{align}
when we take $\theta_0 = iR$.
We have defined
\be
\ell_{ij} \equiv \ell(\HH_i, \HH_j)\,.
\ee

We can define the two-point function on $\HH^3$ as
\be
\label{defH3}
\GG_{k} (\HH,\HH') = {e^{ik \ell(\HH,\HH')} \ov \sinh \ell(\HH,\HH')} \,.
\ee
Unlike in the case of $S^3$ there are two distinct solutions to
\be
(\Box+k_i^2)G_{k_i}(\HH , \HH') ={1 \ov \sqrt{g_{\HH^3}}}  \delta (\HH, \HH')\,.
\ee
In \eq{defH3}, we have chosen a propagator that has ``definite
asymptotic behavior," {\it i.e.,}
\be
\GG_{k} (\ell) \sim e^{(-1+ik) \ell}
\ee
for large $\ell$. With this definition we may write
\begin{align}
\begin{split}
G_{k_j} ((iR_0,\om_0),(iR_j,\om_j))
={e^{ k_j \pi}\GG_{-k_j} (\ell_{j0})-e^{- k_j \pi} \GG_{k_j} (\ell_{j0}) \ov 2i \sinh (k_j \pi) } \,.
\end{split}
\end{align}
Therefore
\begin{align}
\begin{split}
&u^1_\sof{k_i} (\sof{\HH_i}) \\
&={1 \ov 8 \prod_i \sinh k_i \pi} \sum_{\text{signs}}  (\mp 1)(\mp 1)(\mp 1) e^{(\mp k_1 \mp k_2 \mp k_3) \pi}
\int_0^\infty d R_0 \sinh^2 R_0 \int d \om_0 \prod_i  \GG_{\pm k_i} (\ell_{i0}) \\
&= {1 \ov 8 \prod_i \sinh k_i \pi} \sum_{\text{signs}}  (\mp 1)(\mp 1)(\mp 1) e^{(\mp k_1 \mp k_2 \mp k_3) \pi}
\int d \HH_0 \prod_i  \GG_{\pm k_i} (\HH_0,\HH_i) \,,
\end{split}
\end{align}
where the sum are taken over the 8 combinations of assignments of signs.
We can improve the notation by writing
\begin{align}
\begin{split}
&u^1_\sof{k_i} (\sof{\HH_i}) \\
&={1 \ov 8 \prod_i \sinh k_i \pi} \sum_{\sof{\s_i}}  (-\s_1)(-\s_2)(-\s_3) e^{(-\s_1 k_1 -\s_2 k_2 -\s_3 k_3) \pi}
\int_0^\infty d R_0 \sinh^2 R_0 \int d \om_0 \prod_i  \GG_{\s_i k_i} (\ell_{i0}) \\
&= -{1 \ov 8 \prod_i \sinh k_i \pi} \sum_{\sof{\s_i}}   \s_1 \s_2 \s_3 e^{-(\sum_i \s_i k_i) \pi}
\int d \HH_0 \prod_i  \GG_{\s_i k_i} (\HH_0,\HH_i) \,,
\end{split}
\end{align}
where each $\s_i$ runs over the two values $(+1)$ and $(-1)$.

Meanwhile
\be
\UU_\sof{k_i}(\sof{\HH_i}) = \int d \HH_0 \prod_i  \GG_{k_i} (\HH_0,\HH_i)
\ee
is precisely the three-point function on $\HH^3$.
At large values of $R_1=R_2=R_3 =R$, $\UU$ behaves as
\be
\UU_\sof{k_i}(\sof{\HH_i}) \sim \OO(e^{-3R+i(k_1+k_2+k_3)R}) \,.
\ee
Then
\be
u^1_\sof{k_i} (\sof{\HH_i}) 
= -{1 \ov 8 \prod_i \sinh k_i \pi} \sum_{\sof{\s_i}}
\s_1 \s_2 \s_3 e^{-(\sum_i \s_i k_i)  \pi}
\UU_\sof{\s_i k_i} (\sof{\HH_i}) \,.
\ee

Likewise we can do the contour integration along $C_3$
by setting $\theta_0 = iR_0 +\pi$. Using
\begin{align}
\begin{split}
\cos \Th ((iR+\pi,\om),(iR',\om'))
= \cosh (\ell(\HH, \HH')-i\pi) \,,
\end{split}
\end{align}
we find that
\begin{align}
\begin{split}
G_{k_j} ((iR_0,\om_0),(iR_j,\om_j))
={-\GG_{-k_j} (\ell_{j0})+\GG_{k_j} (\ell_{j0}) \ov 2i \sinh (k_j \pi) }
\end{split}
\end{align}
and
\begin{align}
\begin{split}
&u^3_\sof{k_i} (\sof{\HH_i}) \\
&={1 \ov 8 \prod_i \sinh k_i \pi} \sum_{\sof{\s_i}} \s_1 \s_2 \s_3
\int_\infty^0 d R_0 \sinh^2 R_0 \int d \om_0 \prod_i  \GG_{\s_i  k_i} (\ell_{i0}) \\
&= -{1 \ov 8 \prod_i \sinh k_i \pi}  \sum_{\sof{\s_i}} \s_1 \s_2 \s_3
\int d \HH_0 \prod_i  \GG_{\s_i k_i} (\HH_0,\HH_i) \,.
\end{split}
\end{align}

Adding all the contributions up, the analytic continuation
of $U$ becomes
\be
u_\sof{k_i} (\sof{\HH_i}) = 
-{1 \ov 8 \prod_i \sinh k_i \pi} \sum_{\sof{\s_i}} \s_1 \s_2 \s_3
(1+ e^{-(\sum_i \s_i k_i)  \pi})  \UU_\sof{\s_i k_i} (\sof{\HH_i}) \,.
\ee

It is useful to write $\UU_\sof{k_i}$---the three-point function
on $\HH^3$---in Poincar\'e coordinates.
Taking the $\HH^3$ coordinates to be $(z,\vec{x})$ where
$\vec{x} = (x_1,x_2)$ and the metric is given by
\be
ds^2 = {dz^2 + d\vec{x}^2 \ov z^2}\,,
\ee
the two-point correlator satisfies
\be
\GG_{k} (z,\vec{x};z',\vec{x}') = z^{(1-ik)} \left[ {K_{\Delta=1-ik} (\vec{x};z',\vec{x}') \ov -2ik} \right]
\ee
as $z \rightarrow 0$ \cite{KlebanovWitten}.
Using the results of \cite{Freedman},
we find that the three-point correlator
\begin{align}
\begin{split}
\UU_\sof{k_i} (\sof{\HH_i=(z_i,\vec{x}_i)})
=\int {dz d^2\vec{x} \ov z^3} \prod_i \GG_{k_i} (z_i,\vec{x}_i;z,\vec{x})
\end{split}
\end{align}
in the $z_i \rightarrow 0$ limit goes to
\begin{align}
\begin{split}
&\UU_\sof{k_i} (\sof{\HH_i=(z_i,\vec{x}_i)}) \\
&\rightarrow {z_1^{(1-ik_1)}z_2^{(1-ik_2)}z_3^{(1-ik_3)} \ov 16 \pi^2}
{\Ga({\Delta_{12} \ov 2})\Ga({\Delta_{23} \ov 2})\Ga({\Delta_{31} \ov 2})\Ga({1 \ov 2} (\Delta_1+\Delta_2+\Delta_3-2)) \ov
\Ga(\Delta_1)\Ga(\Delta_2)\Ga(\Delta_3)} \\
&\times {1 \ov |x_{12}|^{\Delta_{12}}|x_{23}|^{\Delta_{23}}|x_{31}|^{\Delta_{31}} }
\end{split}
\end{align}
where $\Delta_l = 1-ik_l$, $|x_{lm}|=|\vec{x}_l - \vec{x}_m|$ and
\be
\Delta_{lm} = \Delta_l+\Delta_m -\epsilon_{lmn} \Delta_n
=1-ik_l-ik_m +i\epsilon_{lmn} k_n\,.
\ee
It proves convenient to denote
\be
\tau_\sof{\Delta_i} (\sof{\vec{x}_i} ) \equiv {1 \ov |x_{12}|^{\Delta_{12}}|x_{23}|^{\Delta_{23}}|x_{31}|^{\Delta_{31}} }
\ee
and
\be
c_\sof{\Delta_i} \equiv
{\Ga({\Delta_{12} \ov 2})\Ga({\Delta_{23} \ov 2})\Ga({\Delta_{31} \ov 2})\Ga({1 \ov 2} (\Delta_1+\Delta_2+\Delta_3-2)) \ov
\Ga(\Delta_1)\Ga(\Delta_2)\Ga(\Delta_3)} \,.
\ee

\subsection{The Holographic Series Expansion of the Full Three-point Function}
\label{ss:holexp}

Let us now analytically continue the full three-point function
on the Euclidean instanton \eq{full0}:
\begin{align}
\begin{split}
& T(\sof{X_i},\sof{\Om_i}) \\
&=\int dX W(X)
\int d \Om \prod_i
\left( \int_C {dk_i \ov 2\pi} {\Phi_{k_i} (X_i) \ov \sT(k_i)} \Psi_{k_i}(X) G_{k_i} (\Om, \Om_i)\right) \\
&=\left( \prod_i  \int_C {dk_i \ov 2\pi} {\Phi_{k_i} (X_i) \ov \sT(k_i)}  \right)
\left( \int dX W(X) \prod_i \Psi_{k_i}(X) \right)
U_\sof{k_i} (\sof{\Om_i}) \\
&=\left( \prod_i  \int_C {dk_i \ov 2\pi} {\Phi_{k_i} (X_i) \ov \sT(k_i)}  \right)
S(\sof{k_i}) U_\sof{k_i} (\sof{\Om_i}) \,.
\end{split}
\end{align}
We are interested in the ``near boundary limit" of the correlators.
That is, we look at the behavior of the three-point function as we
take the three points near the boundary of $\HH^3$, {\it i.e.,} we take
$z_i \rightarrow 0$ in Poincar\'e coordinates.

We first analytically continue the spherical coordinates $\Om_j$ to
$\HH_j$ by taking $\theta_j \rightarrow iR_j$.
As elaborated in the previous section, $U_\sof{k_i} (\sof{\Om_i})$ analytically
continues to $u_\sof{k_i} (\sof{\Om_i})$.
By the discussion at the end of the last section we know that
the near boundary limit of $u$ is given by
\begin{align}
\begin{split}
u_\sof{k_i} (\sof{\HH_i}) &=
-{1 \ov 8 \prod_i \sinh k_i \pi} \sum_{\sof{\s_i}} \s_1 \s_2 \s_3
(1+ e^{-(\sum_i \s_i k_i)  \pi})  \UU_\sof{\s_i k_i} (\sof{\HH_i}) \\
&\rightarrow
-{1 \ov 128 \pi^2 \prod_i \sinh k_i \pi} \sum_{\sof{\s_i}} \s_1 \s_2 \s_3
(1+ e^{-(\sum_i \s_i k_i)  \pi})  \\
&\times  (\prod_i z^{1-i \s_i k_i} ) c_\sof{1- i\s_i k_i} \tau_\sof{1-i \s_i k_i} (\sof{\vec{x}_i}) \,.
\end{split}
\end{align}

Now defining
\be
\hS_\sof{\s_i} (\sof{k_i}) \equiv 
{(1+ e^{-(\sum_i \s_i k_i)  \pi}) c_\sof{ 1- i \s_i k_i} \ov 128 \pi^2 \prod_i \sinh k_i \pi}
S(\sof{k_i})
\ee
we find that in the near-boundary limit,
\begin{align}
\begin{split}
& T(\sof{X_i},\sof{\HH_i}) \\
&=-\left( \prod_i  \int_C {dk_i \ov 2\pi} {\Phi_{k_i} (X_i) \ov \sT(k_i)}  \right)
\sum_\sof{\s_i} \s_1 \s_2 \s_3 \hS_\sof{\s_i}(\sof{k_i}) (\prod_i z^{1- i\s_i k_i} )  \tau_\sof{1- i\s_i k_i} (\sof{\vec{x}_i}) \,
\end{split}
\end{align}
as $z \rightarrow 0$.

In this limit, the $|k_i| \rightarrow \infty$ behavior is dictated by $z^{1\mp ik_i}$.
That is, for terms of the integrand with $z$ dependence $z^{1- ik_i}$
we may deform the contour $C$ ``upwards"
towards $k_i \rightarrow i \infty$,
while for terms with $z$ dependence $z^{1+ ik_i}$
we can deform the contour $C$ ``downwards"
towards $k_i \rightarrow -i \infty$.
More concisely, we may deform terms with $z$ dependence $z^{1-i\s_i k_i}$
towards $k_i \ra i \s_i \infty$ when $z \ra 0$.
By deforming the terms in the integrand accordingly,
the integral can be written as a sum of contributions from
codimension-three poles of the integrand, {\it i.e.,} poles of the form
\be
E^\sof{\sigma_i}_\sof{k_i} (\sof{X_i})
\equiv \left( \prod_i {\Phi_{k_i} \ov \sT(k_i) } \right)
\hS_\sof{\s_i}(\sof{k_i}) \sim {x \ov (k_1-p_1)(k_2-p_2)(k_3-p_3)} + \cdots
\ee
near $\sof{k_i} = \sof{p_i}$.

%In the event that $(\Phi_{k_i}/ \sT(k_i))$ does not have any poles,
%a simplification occurs. In this case all we need to keep track are
%poles of $\hS_\sof{\s_i}(\sof{k_i})$.
%This is true in the important case that the scalar is massless in the
%flat region.
%Through the rest of this subsection we assume that
%$(\Phi_{k_i}/ \sT(k_i))$ has no poles and comment on the case
%when it does have poles at the end.

Let us denote the $H_i^{(+1)}$ the upper-half of the complex $k_i$ plane divided by
contour $C$ and $H_i^{(-1)}$ the lower-half.
Let us also denote the codimension-three poles of $\hS_\sof{\s_i} (\sof{k_i})$
with respect to $k_i$ in $H_1^{\s_1} \times H_2^{\s_2} \times H_3^{\s_3}$ as 
$(i\s_1 n_{r_1}^{\s_1},i\s_2 n_{r_2}^{\s_2}, i\s_3 n_{r_3}^{\s_3})$,
or in short, $\sof{i\s_i n_{r_i}^{\s_i}}$.
Then by doing the contour integral and picking up the poles of the integrand
for each pole we obtain
\begin{align}
\begin{split}
& T(\sof{X_i},\sof{\HH_i}) \\
&=-\left( \prod_i  \int_C {dk_i \ov 2\pi} {\Phi_{k_i} (X_i) \ov \sT(k_i)}  \right)
\sum_\sof{\s_i} \s_1\s_2\s_3 \hS_\sof{\s_i}(\sof{k_i}) (\prod_i z^{1- i\s_ik_i} )  \tau_\sof{1- i\s_i k_i} (\sof{\vec{x}_i}) \\
&=  i \sum_\sof{\s_i} \sum_\sof{n_{r_i}^{\s_i}} 
\hat{F}^\sof{\s_i}_\sof{n^{\s_i}_{r_i}} (\sof{X_i}) ( \prod_i z^{n^{\s_i}_{r_i}+1} )
 \tau_\sof{n^{\s_i}_{r_i}+1} (\sof{\vec{x}_i}) \\
 &\quad + \text{(Double Pole Contributions)}
\end{split}
\label{preresult}
\end{align}
where we have defined
\be
\hat{F}^\sof{\s_i}_\sof{n^{\s_i}_{r_i}} (\sof{X_i})
= \Res{\sof{k_i}=\sof{i\s_i n^{\s_i}_{r_i}}} E^\sof{\sigma_i}_\sof{k_i} (\sof{X_i})\,.
\ee
As indicated in the equation there may be additional terms coming from the double
poles of $E^\sof{\sigma_i}_\sof{k_i} (\sof{X_i})$ with respect to $k_i$,
which certainly exist in general.

Finally we analytically continue $X_i$ to obtain the holographic
expansion of the three-point function:
\[
\fbox{
\addtolength{\linewidth}{-2\fboxsep}%
\addtolength{\linewidth}{-2\fboxrule}%
\begin{minipage}{\linewidth}
\begin{align}
\begin{split}
& \langle \vh(T_1,\HH_1)\vh(T_2,\HH_2)\vh(T_3,\HH_3) \rangle \\
&= \left[ \prod_i a(T_i-i\pi/2)^{-1} \right] T(\sof{T_i-i\pi/2},\sof{\HH_i}) \\
&= (-i) \left[ \prod_i \ta(T_i)^{-1} \right] T(\sof{T_i-i\pi/2},\sof{\HH_i}) \\
&= \sum_\sof{\s_i} \sum_\sof{n_{r_i}^{\s_i}} 
F^\sof{\s_i}_\sof{n^{\s_i}_{r_i}+1} (\sof{T_i}) ( \prod_i  z^{n^{\s_i}_{r_i}+1} )
 \tau_\sof{n^{\s_i}_{r_i}+1} (\sof{\vec{x}_i}) \\
& \qquad+ \text{(Double Pole Contributions)}
 \label{holexp}
\end{split}
\end{align}
\end{minipage}\nonumber
}
\]
$F^\sof{\s_i}_\sof{n^{\s_i}_{r_i}+1} (\sof{T_i})$ is defined as
\be
F^\sof{\s_i}_\sof{n^{\s_i}_{r_i}+1}(\sof{T_i})
\equiv
\left[ \prod_i \ta(T_i)^{-1} \right]
\Res{\sof{k_i}=\sof{i\s_i n^{\s_i}_{r_i}}}
E^\sof{\sigma_i}_\sof{k_i} (\sof{T_i-i\pi/2}) \,.
\ee

Now let us examine the behavior of $F^\sof{\s_i}_\sof{n_i+1} (\sof{T_i})$
at early times.
$\Phi_k/\sT(k)$ has exponential $T$ scaling at early times.
More precisely, $\Phi_k(X)/\sT(k)$ can be written as
\be
{\Phi_k(X) \ov \sT(k)} = e^{-ikX} \left( 1+ \sum_{n=1}^\infty c_n (k) e^{nX} \right)
\ee
near $X \ra -\infty$.
$c_n$ may have poles in $k$. From the definition of 
$E^\sof{\sigma_i}_\sof{k_i} (\sof{X_i})$
it can be shown that in the early-time limit
\begin{align}
\begin{split}
&\left[ \prod_i \ta(T_i)^{-1} \right] E^\sof{\sigma_i}_\sof{k_i} (\sof{T_i-i\pi/2}) \\
&=  \left( \prod_i {\Phi_{k_i} (T_i-i\pi/2)\ov \ta(T_i) \sT(k_i) } \right) \hS_\sof{\s_i} (\sof{k_i}) \\
&\ra \left({1 \ov L^3} \right)  \hS_\sof{\s_i} (\sof{k_i})
e^{-\sum_i k_i \pi/2 } e^{\sum_i (- 1 -ik_i )T_i} \,.
\end{split}
\end{align}
The terms in \eq{holexp} relevant in the early-time limit $T_i \ra -\infty$
are given by poles of $\hS_\sof{\s_i} (\sof{k_i})$ when $\sof{\s_i}=(-1,-1,-1)$---they
come from picking up poles of the integrand that are in the lower-half $k$ plane.
The terms of the holographic expansion coming from these poles
are proportional to
\be
( \prod_i  e^{-\Delta_i T_i} z^{\Delta_i} )
\tau_\sof{\Delta_i} (\sof{\vec{x}_i})
\label{confform}
\ee
for $\Delta_i = n^{-}_{r_i} +1$,
as promised in the introduction.

In general, terms in the expansion that come from double poles
have factors of $(\ln T)$, $(\ln z)$ or $(\ln |x_{ij}|)$ multiplied to the
``conformal" form \eq{confform} in the early-time limit.
It can be seen, however, that any double pole
coming from poles of $\Phi_k /\sT(k)$ are due to poles of $c_n$.
Hence double pole contributions of this kind are always
subleading in $e^T$ in the early-time limit and can be ignored.

To write out all the terms of the early-time holographic expansion,
one must locate all the poles of the integrand carefully and compute its residues
and take the early-time limit.
There are many terms that have to be computed on a case-by-case basis
that depend on the structure of $\Phi_k/\sT(k)$ and $S(\sof{k_i})$.
Most of the terms, however, are ``universal" in any early-time holographic
expansion. These terms come from picking up ``generic poles"
at $\sof{-in^{-}_{r_i}} \in H^{-1}_{k_1} \times H^{-1}_{k_2} \times H^{-1}_{k_3}$.
We define ``generic poles"
$\sof{i\s_i n^{\s_i}_{r_i}} \in H^{\s_1}_{k_1} \times H^{\s_2}_{k_2} \times H^{\s_3}_{k_3}$
to be points that satisfy the following conditions:
\begin{enumerate}
\item $n^{\s_i}_{r_i}$ are integers.
\item The triplet $\sof{n^{\s_i}_{r_i}+1}$ satisfies the triangle inequality.
\item $\sof{k_i} = \sof{ i\s_i  n^{\s_i}_{r_i}}$ is either
\begin{enumerate}
\item not a pole of $S(\sof{k_i})$,
\item or lies on a codimension-one pole of $S(\sof{k_i})$
and $\sum_i n^{\s_i}_{r_i}$ is odd.
\end{enumerate}
\end{enumerate}
The contribution of a generic pole $\sof{k_i} = \sof{-i n^{-}_{r_i}}$
to the early-time three-point function is
\[
\fbox{
\addtolength{\linewidth}{-2\fboxsep}%
\addtolength{\linewidth}{-2\fboxrule}%
\begin{minipage}{\linewidth}
\begin{align}
\begin{split}
&\ra \left( {i \ov  64\pi^5 L^3} \right)
e^{-i\pi(\sum_i \Delta^-_{r_i})/2}\sin \left( {\pi \sum_i \Delta^-_{r_i} \ov 2} \right)
{S(\sof{- i(\Delta_{r_i}^- -1)})}c_\sof{\Delta^{-}_{r_i}} \\
&\qquad \times ( \prod_i  e^{-\Delta^{-}_{r_i} T_i} z^{\Delta^{-}_{r_i}} )
 \tau_\sof{\Delta^{-}_{r_i}} (\sof{\vec{x}_i})
\end{split}
\end{align}
\vskip2pt
\end{minipage}\nonumber
}
\]
where we have defined $\Delta^{\s}_{r} = n^{\s}_r+1$.

We note that contrary to the early-time limit $T_i \rightarrow -\infty$, we cannot
obtain the late time limit $T_i \rightarrow \infty$ of $\Phi_k$ by plugging in
$X \rightarrow T-i\pi/2$
to the $X \rightarrow \infty$ limit of $\Phi_k$ in \eq{Philim}.
This is because we are ``gluing" the modes on the Euclidean instanton
and the cosmological background at $r= 0$, where $r$ is the
radial coordinate of the instanton. This is equivalent
to gluing the modes at
$X, T \rightarrow -\infty$ with respect to the conformal coordinates
$X$ and $T$.

\subsection{Spherical Harmonics expansion of the Three-Point function\\
on the Past Lightcone}
\label{ss:harm}

Let us expand the three-point function in terms of $S^2$ harmonics
when the points lie on the past lightcone of the FRW region, {\it i.e.,} when 
\begin{align}
T_1, T_2, T_3 &\rightarrow -\infty \\
R_1, R_2, R_3 &\rightarrow \infty \\
T_i +R_i  &: \text{(finite)}
\end{align}

To do so, let us go back to expression \eq{full0} and analytically
continue the $S^3$ coordinates to $\HH^3$ coordinates:
\begin{align}
\begin{split}
&T(\sof{X_i},\sof{\HH_i}) \\
&=-\left( \prod_i  \int_C {dk_i \ov 4\pi \sinh k_i \pi} {\Phi_{k_i} (X_i) \ov \sT(k_i)}  \right)
S(\sof{k_i}) \\
&\quad \times \sum_\sof{\s_i} \s_1\s_2\s_3
(1+ e^{-(\sum_i \s_i k_i) \pi})  \UU_\sof{\s_i k_i} (\sof{\HH_i}) 
\,.
\end{split}
\end{align}
Now let us write $\UU_\sof{k_i}$ in terms of harmonics.
Recall that $\UU_\sof{k_i}$ is defined by
\be
\UU_\sof{k_i}(\sof{\HH_i}) = \int d \HH_0 \prod_i  \GG_{k_i} (\HH_0,\HH_i)\,.
\ee
The correlators $ \GG_{k} (\HH_0,\HH)$ can be written in terms
of scalar eigenmodes on $\HH^3$ \cite{CamporesiHiguchi,STY}
\be
q_{klm} (\HH) \equiv q_{kl} (R) Y_{lm} (\om) \,,
\ee
where
\begin{align}
\begin{split}
q_{kl} (R) &= (i \sinh R)^l F(ik+l+1,-ik+l+1;l+{3 \ov 2};-\sinh^2 {R \ov 2}) \\
&= {2(-2i)^{l} \Ga(l+3/2) \ov \sqrt{\pi} k\prod_{j=1}^l (k^2+j^2)} (\sinh R)^l
({d \ov d \cosh R})^l {\sin kR \ov \sinh R} \,,
\end{split}
\end{align}
and $Y_{lm}$ are the spherical harmonic functions.

\begin{figure}[!t]
\begin{minipage}[b]{0.5\linewidth}
\centering\includegraphics[width=6cm]{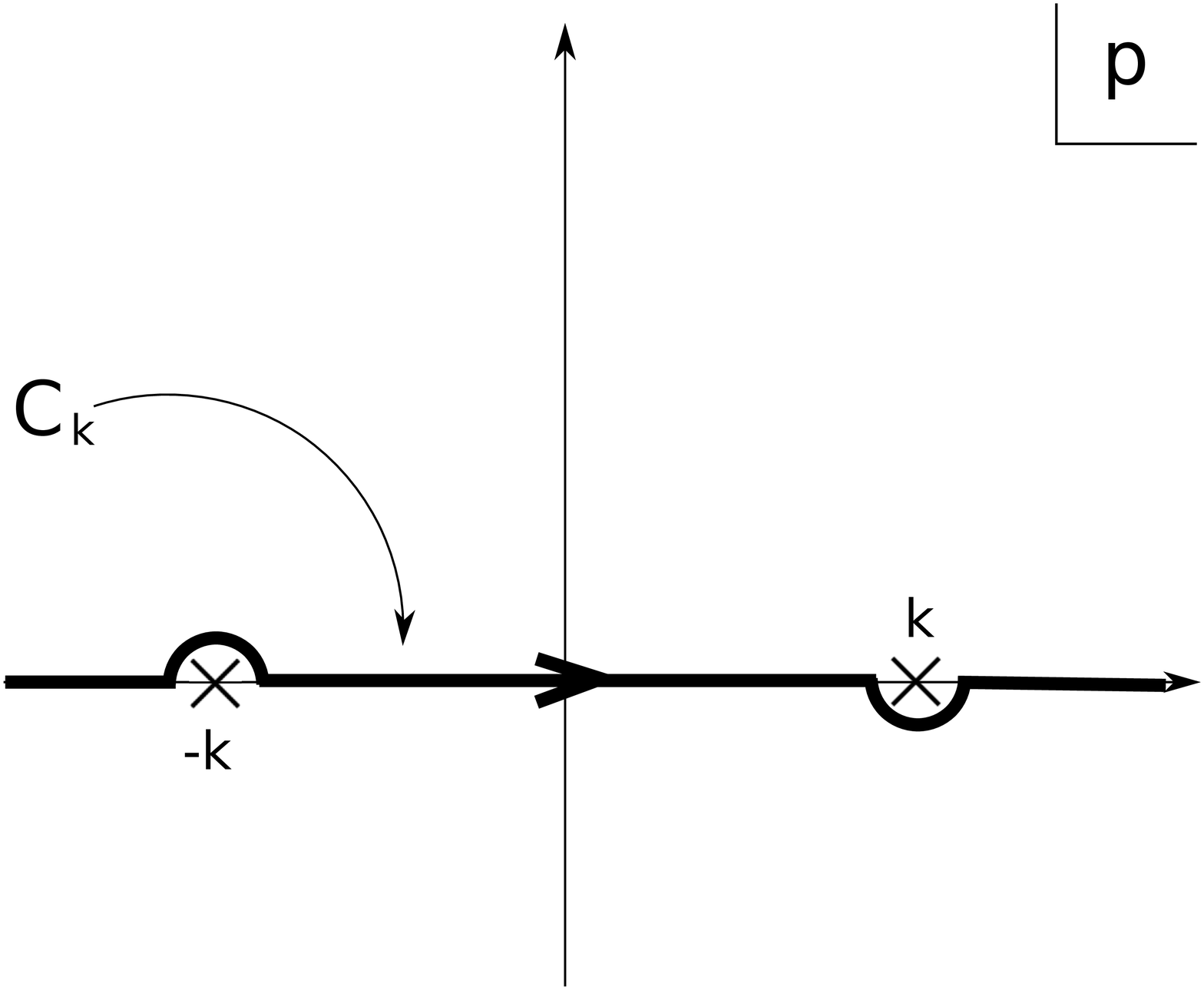}
\end{minipage}%
\begin{minipage}[b]{0.5\linewidth}
\centering\includegraphics[width=6cm]{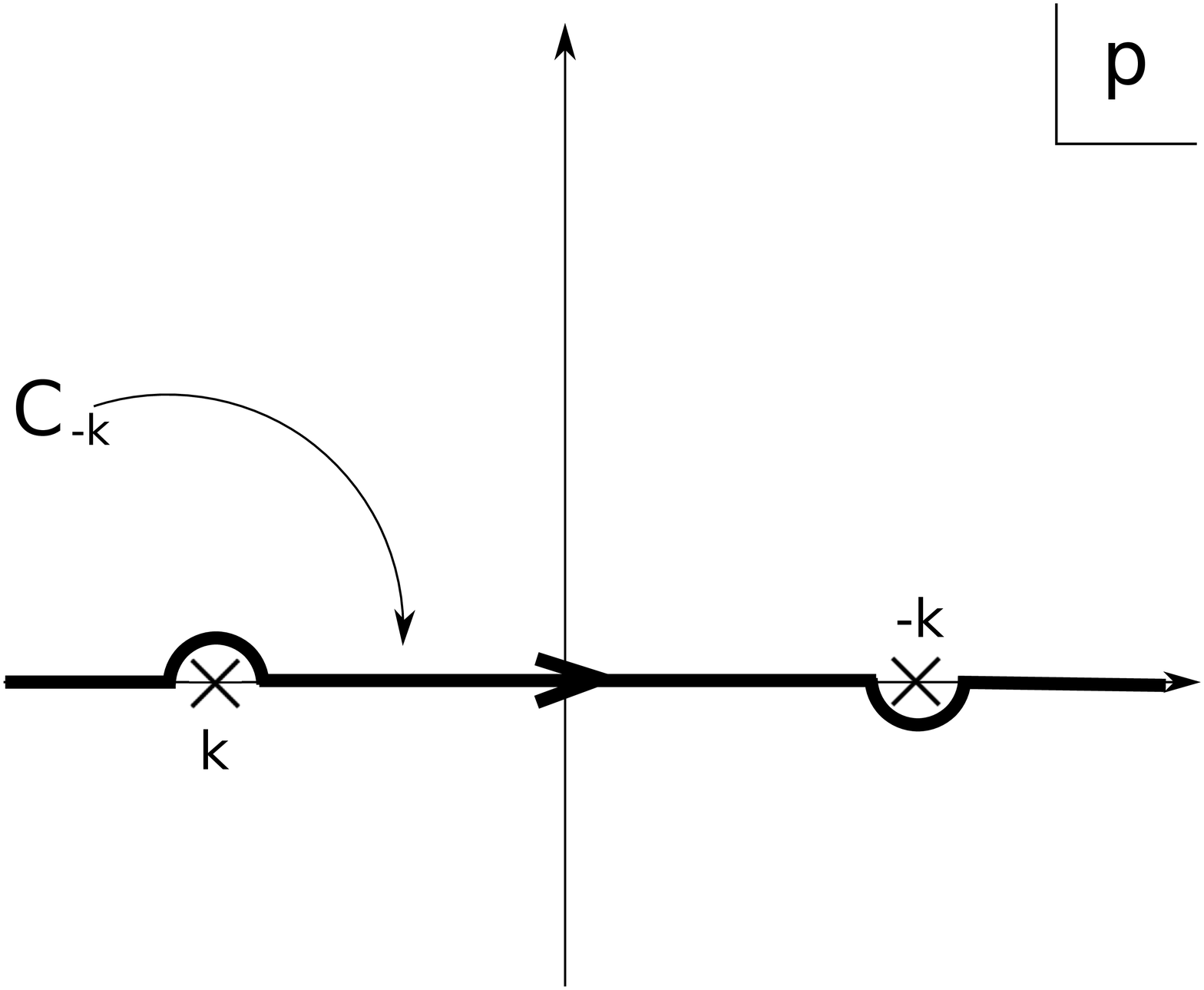}
\end{minipage}
\caption{\small The contour $C_k$. The contour
$C_{-k}$ follows from the definition of $C_{k}$.}
\label{f:ck}
\end{figure}

Then
\be
 \GG_{k} (\HH_0,\HH)
 ={1 \ov 2}\sum_{l,m} \int_{C_k} dp {N_{l} (p) \ov p^2 -k^2} q_{pl(-m)} (\HH_0 ) q_{plm} (\HH)  \,,
\ee
where $C_k$---depicted in figure \ref{f:ck}---is the contour of integration.
$N_{l} (p)$ is a normalization constant:
\begin{align}
\begin{split}
N_{l} (p) &= {(-1)^l  \Ga(ip+l+1) \Ga(-ip+l+1) \ov 2^{2l+1}
\Ga(l + {3 \ov 2})^2  \Ga(ip) \Ga(-ip)}
= {(-1)^l\prod_{j=0}^l (p^2 + j^2) \ov 2^{2l+1} \Ga(l+3/2)^2}
\,.
\end{split}
\end{align}

Then $\UU_\sof{k_i}$ can be written as
\begin{align}
\begin{split}
&\UU_\sof{k_i} (\sof{\HH_i})=   \\
&{1 \ov 8}\sum_{l_i,m_i} 
\left[ \prod_i \left( \int_{C_{k_i}} {dp_i \ov p_i^2 -k_i^2 } \right) 
\left(  \prod_i q_{p_i l_i m_i} (\HH) \right)   
\int d \HH_0 \left( \prod_i {N_{l_i} (p_i) q_{p_i l_i (-m_i)} (\HH_0 ) } \right)  \right] \,.
\end{split}
\end{align}
We define
\be
B(\sof{p_i,l_i,m_i}) \equiv \int d \HH_0 \left( \prod_i {N_{l_i} (p_i) q_{p_i l_i (-m_i)} (\HH_0 ) } \right) 
\label{defB}
\ee
which can be thought of as structure functions on $\HH^3$
similar to Wigner coefficients on $S^2$.
We have listed some important properties of $B(\sof{p_i,l_i,m_i})$
in appendix \ref{ap:b}.

\begin{figure}[!t]
\begin{minipage}[b]{0.5\linewidth}
\centering\includegraphics[width=6cm]{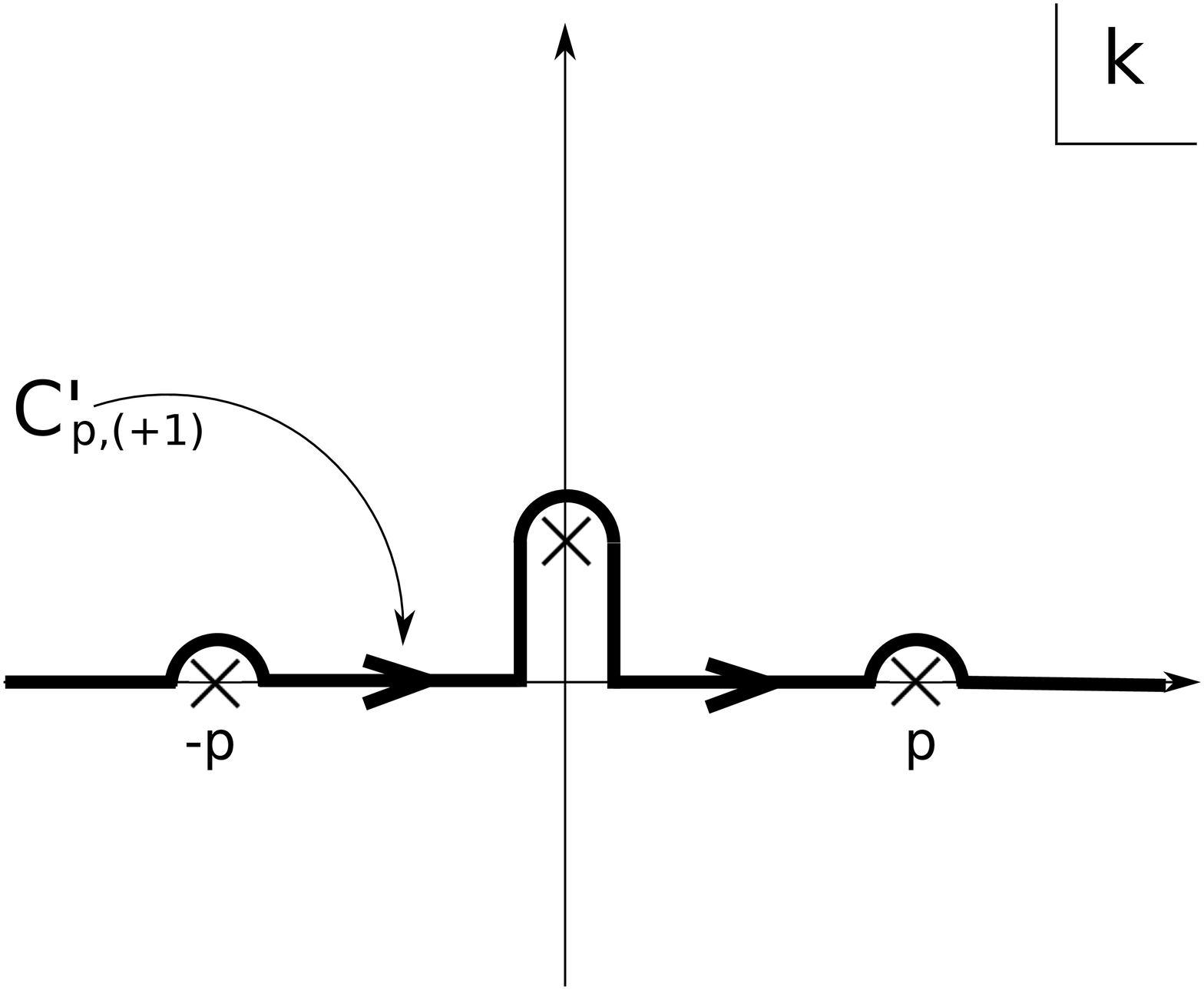}
\end{minipage}%
\begin{minipage}[b]{0.5\linewidth}
\centering\includegraphics[width=6cm]{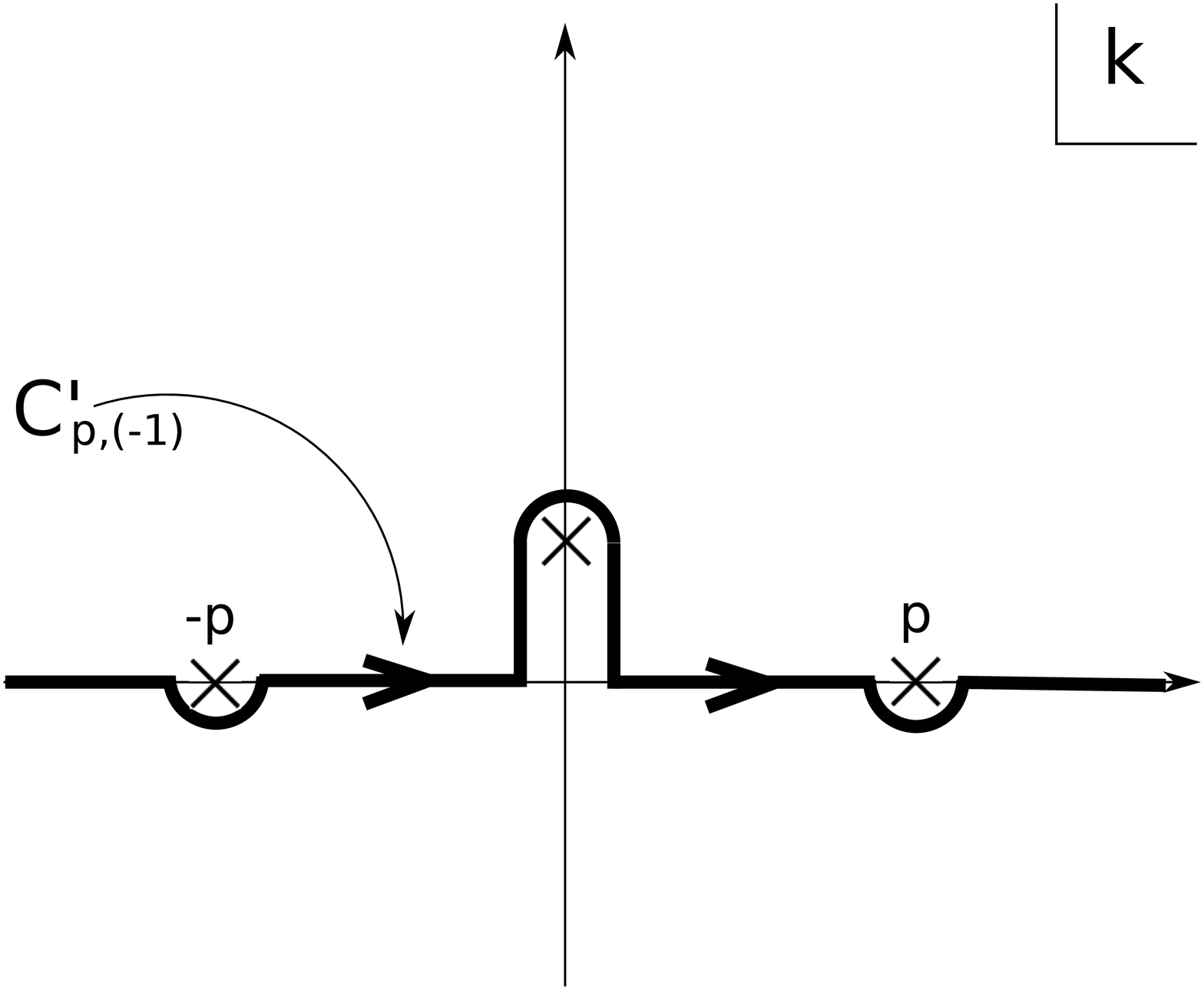}
\end{minipage}
\caption{\small The contours $C'_{p,(+1)}$ and $C'_{p,(-1)}$.
The cross on the imaginary axis is the upmost pole of $\sT(k)$.}
\label{f:cp}
\end{figure}

The analytically continued full three-point function may be rewritten as
\begin{align}
\begin{split}
& T(\sof{X_i},\sof{\HH_i}) \\
&=-{1 \ov 64} \sum_\sof{\s_i}  \prod_i \left( \int_{-\infty}^\infty dp_i \right)
\sum_{l_i,m_i} B(\sof{p_i,l_i,m_i})
 \left( \prod_i q_{p_i l_i m_i} (\HH_i) \right) \\
&\quad \left( \prod_i  \int_{C'_{p_i,\s_i }} {dk_i \Phi_{k_i} (X_i) \ov 2\pi \sT(k_i) \sinh k_i \pi (p_i^2 -k_i^2)}  \right)
S(\sof{k_i}) \s_1\s_2\s_3
(1+ e^{-(\sum_i \s_i k_i) \pi})
 \,.
\end{split}
\end{align}
As we have switched the order of integration,
the contour of integration of the $k_i$'s have been changed
to $C'_{p_i,\s_i}$ which is depicted in figure \ref{f:cp}.

In the $X\rightarrow (-\infty -i\pi/2)$ limit, or the $T \rightarrow -\infty$ limit,
there is a useful expansion in terms of $e^{T}$.
In this limit
\be
{\Phi_k (T-i\pi/2) \ov \sT (k)} \rightarrow e^{-ik(T-i\pi/2)}
\ee
Therefore all the contours of integration for $k_i$ can be deformed upward
and the poles $k_i$ with $({\rm Im}~ k_i) > 0$
and possibly $k_i = \pm p_i$ contribute.
Hence
\begin{align}
\begin{split}
&- \sum_\sof{\s_i} \left( \prod_i  \int_{C'_{p_i,\s_i}} {dk_i e^{-ik_i (T_i-i\pi/2)} \ov 2\pi \sinh k_i \pi (p_i^2 -k_i^2)}  \right)
S(\sof{k_i}) \s_1\s_2\s_3
(1+ e^{-(\sum_i \s_i k_i) \pi}) \\
&= {i \ov 4 \prod_i p_i \sinh \pi p_i} \sum_\sof{s_i}
S(\sof{s_i p_i})  \cosh {\pi (\sum_i s_i p_i) \ov 2 }e^{-i \sum_i s_i p_i T_i} \\
&+ \sum_{\sof{\s_i} \neq (-1,-1,-1)} \sum_{(\text{at least one }n_i>0)}
\OO( e^{\sum_i n_i T_i})
 \,.
\end{split}
\end{align}
The leading term comes from the poles $\pm p_i$ that are picked up
for the contour integral along $C_{k_1,(-1)} \times C_{k_2,(-1)} \times C_{k_3,(-1)}$.
Each $s_i$ is summed over $(+1)$ and $(-1)$. 
The subleading terms in $e^T$ come from other combinations
of contours.

In the limit  $T_1,T_2,T_3 \rightarrow -\infty$ the integrand is equal to
the leading term. Therefore on the past lightcone
\begin{align}
\begin{split}
& T(\sof{T_i-i\pi/2},\sof{\HH_i}) \\
&= {1 \ov 64} \prod_i \left( \int_{-\infty}^\infty dp_i \right)
\sum_{l_i,m_i} B(\sof{p_i,l_i,m_i})
 \left( \prod_i  q_{p_i l_i m_i} (\HH_i) \right) \\
&\quad \times  {i \ov 4 \prod_i p_i \sinh \pi p_i} \sum_\sof{s_i}
S(\sof{s_i p_i})  \cosh {\pi (\sum_i s_i p_i) \ov 2 }e^{-i \sum_i s_i p_i T_i}  \,.
\end{split}
\end{align}
$B(\sof{p_i,l_i,m_i})$, $q_{p_i l_i (-m_i)} (\HH_i)$
and $p_i \sinh \pi p_i$ are all even with respect to $p_i$.
Therefore we can rewrite the integral as
\begin{align}
\begin{split}
& T(\sof{T_i-i\pi/2},\sof{\HH_i}) \\
&= {i \ov 32} \prod_i \left( \int_{-\infty}^\infty dp_i \right)
\sum_{l_i,m_i} B(\sof{p_i,l_i,m_i})
 \left( \prod_i  q_{p_i l_i m_i} (\HH_i) \right) \\
&\quad \times {\cosh{\pi(p_1+p_2+p_3) \ov 2} \ov \prod_i p_i \sinh \pi p_i} 
S(\sof{-p_i}) e^{ip_1 T_1}e^{ip_2 T_2}e^{ip_3 T_3}
\end{split}
\end{align}
for points on the past lightcone.
In appendix \ref{ap:b} we show that $B(\sof{p_i,l_i,m_i})$ has a
zero of order two for each $p_i$ at $p_i =0$.
Hence the integrand does not have any poles along the real line.

In the large $R$ limit, $q_{plm} (R,\om) $ behaves as
\cite{CamporesiHiguchi}
\begin{align}
\begin{split}
q_{plm} (R,\om)
&\sim (-i){(2i)^{l+1} \Ga(l+{3 \ov 2}) \Ga(ip) \ov \sqrt{\pi} \Ga(ip+l+1)} e^{(-1+ip)R} Y_{lm} (\om)\\
&+ (-i){(2i)^{l+1} \Ga(l+{3 \ov 2}) \Ga(-ip) \ov \sqrt{\pi} \Ga(-ip+l+1)} e^{(-1-ip)R} Y_{lm} (\om) \,.
\end{split}
\end{align}
Therefore on the past lightcone
\begin{align}
\begin{split}
& T(\sof{T_i-i\pi/2},\sof{\HH_i}) \\
&= \sum_\sof{l_i,m_i} e^{-R_1-R_2-R_3} K_\sof{l_i,m_i}(\sof{T_i,R_i})
Y_{l_1 m_1} (\om_1) Y_{l_2 m_2} (\om_2) Y_{l_3 m_3} (\om_3)
\end{split}
\end{align}
where
\begin{align}
\begin{split}
&K_\sof{l_i,m_i}(\sof{T_i,R_i}) \\
&= -{1 \ov 32}  \prod_i \left( \int_{-\infty}^\infty dp_i \right)
{B(\sof{p_i,l_i,m_i}) \ov \prod_i p_i \sinh \pi p_i} {S(\sof{-p_i})} \cosh{\pi(p_1+p_2+p_3) \ov 2}  \\
&\quad \times  \sum_\sof{\s_i} \left( \prod_i  {(2i)^{l_i+1} \Ga(l_i+{3 \ov 2}) \Ga(i\s_i p_i) \ov \sqrt{\pi} \Ga(i\s_i p_i+l_i+1)}
e^{(i(T_i+\s_i R_i))p_i} \right)
\end{split}
\end{align}

We can deform the contour of integration for terms with factors of $e^{i(T_i -R_i)p_i}$
downward and pick up poles of the integrand to give terms of the form
\be
\sum_r c_r e^{(T-R)n_r}
\ee
where the real part of the $n_r$ are positive.
These terms can be ignored on the past lightcone.
Therefore the relevant term becomes the term with all $\s_i$ being $(+1)$:
\begin{align}
\begin{split}
&K_\sof{l_i,m_i}(\sof{T_i,R_i}) \\
&= -{1 \ov 32}  \prod_i \left( \int_{-\infty}^\infty dp_i \right)
{B(\sof{p_i,l_i,m_i}) \ov \prod_i p_i \sinh \pi p_i} {S(\sof{-p_i})} \cosh{\pi(p_1+p_2+p_3) \ov 2}  \\
&\quad \times\left( \prod_i  {(2i)^{l_i+1} \Ga(l_i+{3 \ov 2}) \Ga(i p_i) \ov \sqrt{\pi} \Ga(i p_i+l_i+1)}
e^{(i(T_i+R_i))p_i} \right)
\end{split}
\end{align}

Finally we obtain
\[
\fbox{
\addtolength{\linewidth}{-2\fboxsep}%
\addtolength{\linewidth}{-2\fboxrule}%
\begin{minipage}{\linewidth}
\begin{align}
\begin{split}
& \langle \vh(T_1,\HH_1)\vh(T_2,\HH_2)\vh(T_3,\HH_3) \rangle \\
&=(-i)\left[\prod_i \ta(T)^{-1}\right]
T((T_1\!-\!i{\pi \ov2},T_2\!-\!i{\pi \ov2},T_3\!-\!i{\pi \ov2}),
((R_1,\om_1),(R_2,\om_2),(R_3,\om_3))) \\
&= \left( {1 \ov L^3} \right)
\prod_i e^{-(T_i+R_i)} \sum_{l_i,m_i} K'_\sof{l_i,m_i} (\sof{T_i+R_i})
Y_{l_1 m_1} (\om_1) Y_{l_2 m_2} (\om_2) Y_{l_3 m_3} (\om_3)
 \,.
\end{split}
\label{harmexp}
\end{align}
\end{minipage}\nonumber
}
\]
where
\[
\fbox{
\addtolength{\linewidth}{-2\fboxsep}%
\addtolength{\linewidth}{-2\fboxrule}%
\begin{minipage}{\linewidth}
\begin{align}
\begin{split}
&K'_\sof{l_i,m_i}(\sof{T_i+R_i}) \\
&= {1 \ov 4}  \prod_i \left( \int_{-\infty}^\infty dp_i e^{i(T_i+R_i)p_i} \right) {S(\sof{-p_i})} \\
&\quad \times  {B(\sof{p_i,l_i,m_i}) \ov \prod_i p_i \sinh \pi p_i}  \cosh{\pi(p_1+p_2+p_3) \ov 2} 
 \left( \prod_i  {(2i)^{l_i} \Ga(l_i+{3 \ov 2}) \Ga( ip_i) \ov \sqrt{\pi} \Ga(ip_i+l_i+1)}
\right)
\,.
\end{split}
\label{harmcoeff}
\end{align}
\end{minipage}\nonumber
}
\]

We note that this expression is valid regardless of the pole
structure of $\Phi_k /\sT(k)$ in the lower-half plane of $k$.
Although we have replaced
\be
{\Phi_k \ov \sT(k)} \rightarrow e^{-ik(T-i\pi/2)}
\ee 
throughout the calculation, we can keep $(\Phi_k /\sT(k))$
and check that all poles picked up by contour deformation
are indeed subleading in $e^T$ and can safely be ignored,
due to regularity.

One might worry that the integrand is not well defined at $p_i=0$ since
\be
 {\Ga( ip_i) \ov \Ga(ip_i+l_i+1)} = {1 \ov ip_i (ip_i+1) \cdots (ip_i+l_i)} \,.
\ee
Since $B(\sof{p_i,l_i,m_i})$ has a
zero of order two for each $p_i$ at $p_i =0$,
we find that the factor in the third line of equation \eq{harmcoeff}
has a simple pole at $p_i=0$ with respect to each $p_i$.
It is the behavior of $S(\sof{p_i})$ near $p_i=0$
that keeps the integrand well-defined.
For generic $U(X)$,
\be
\sR (p) \ra -1, \quad \sT(p) \ra 0 \quad
\text{for }p\ra 0.\footnote{This behavior holds except in special cases
when there is a discrete eigenstate of $U(X)$ at zero energy
\cite{Barton}.}
\label{0momentum}
\ee
This implies that
\be
\Psi_p (X) \ra 0 \quad \text{as }p\ra 0.
\ee
Hence
\be
S(\sof{p_i}) = \int_{-\infty}^\infty dX W(X)
 \Psi_{p_1} (X) \Psi_{p_2} (X) \Psi_{p_3} (X)
\ee has a
zero with respect to each $p_i$ at $p_i =0$,
and the integral is well defined.
One can easily check that the reflection/transmission
coefficients of the eigenmodes of $U(X)$ indeed behave as
\eq{0momentum} for the massless and massive scalars
we study in this paper.

The equation \eq{harmcoeff} implies that the coefficients
of the harmonic expansion of the three-point function can be obtained
by a weighted Fourier transform from the wavefunction overlap.
We expect the structure function multiplying $S(\sof{p_i})$ to have
exponential decay to render the integral well-defined.
For example, when
\be
\sof{l_i,m_i} = ((0,0),(0,0),(0,0))\,,
\ee
a short calculation reveals that $B(\sof{p_i,l_i,m_i}) $
is given by
\begin{align}
\begin{split}
 { ( p_1 p_2 p_3)
\sinh \pi p_1 \sinh \pi p_2 \sinh \pi p_3
\ov 2 \pi^{5/2} \cosh {\pi (-p_1 + p_2 + p_3) \ov 2}
\cosh {\pi (p_1 - p_2 + p_3) \ov 2}
\cosh {\pi (p_1 + p_2 - p_3) \ov 2}
\cosh {\pi (p_1 + p_2 + p_3)\ov 2} } \,.
\end{split}
\end{align}
Therefore
\begin{align}
\begin{split}
&K'_{((0,0),(0,0),(0,0))}(\sof{T_i+R_i}) \\
&= {i \ov 64 \pi^{5/2}}  \prod_i \left( \int_{-\infty}^\infty dp_i e^{i(T_i+R_i)p_i} \right) {S(\sof{p_i})} \\
&\quad \times   { 1 \ov p_1 p_2 p_3 \cosh {\pi (-p_1 + p_2 + p_3) \ov 2}
\cosh {\pi (p_1 - p_2 + p_3) \ov 2}
\cosh {\pi (p_1 + p_2 - p_3) \ov 2} }
\,.
\end{split}
\end{align}

\section{The Massless Scalar in the Thin-wall Limit} \label{s:m0thinwall}

In this section, we compute the three-point function for a specific example.
We set the gravitational background to be a thin-wall CDL instanton which is
flat on one side $(X<X_0)$ and de Sitter on the other $(X>X_0)$.
We assume the potential for the scalar $\vh$ has the expansion
\be
\mathcal{V}(\vh,X) =
\begin{cases}
0+\OO(\vh^4) & (X<X_0) \\
{1 \ov 2} m^2 \vh^2 + {1 \ov 6} \lambda \vh^3 +\OO(\vh^4) & (X>X_0)
\end{cases}
\ee
around the given background.
Such a potential can be obtained by a potential such as
\eq{pot1}.
Note that the scalar is massless on the flat side.

The radial potential $U(X)$ derived from this potential
is regular for small enough $m$, as shown in appendix \ref{aps:m0},
and hence we can use the results of the previous sections.
As can be seen in section \ref{s:ancon},
the data needed in determining three-point function are:
\begin{enumerate}
\item The eigenmodes $\Phi_k, \Psi_k$ of the radial potential.
\item The wavefunction overlap $S(\sof{k_i})$ in the radial direction.
\end{enumerate}
We present these data for our example.

We first review the thin-wall instanton in section \ref{ss:thinwall}.
We present the eigenmodes $\Phi_k, \Psi_k$
and their analytic continuation in section \ref{ss:modes}.
${\Phi_k /\sT(k)}$ turns out to be an exponential
function in $T$. This implies that the three-point function
has a holographic expansion with exponential $T$ scaling \eq{holform}
at all times in the FRW patch---we write the holographic expansion
explicitly in this section.
We compute $S(\sof{k_i})$ in section \ref{ss:overlap}.

\subsection{The Thin-wall CDL Instanton} \label{ss:thinwall}

We define the thin-wall CDL instanton so that there exists
two distinct regions---separated by a thin wall---of the instanton where
the scalar field takes two discrete values $\phi_-$, $\phi_+$
at two distinct local minima of the potential $V$.
We are interested in the case when
$V(\phi_+)>V(\phi_-)=0$.

Then, the thin-wall CDL instanton can be defined as an analytic
continuation of the metric
\begin{align}
 ds^2 &= a^2 (X) (dX^2 + d \Om^2 ) \\
 &= a^2 (X) (dX^2 + d \theta^2 + \sin^2 \theta d \om^2 ) \,,
\end{align}
where we define
\begin{align}
 a(X) = \begin{cases}
      \frac{e^{X-X_0 }}{\cosh X_0 } & (X < X_0) \\
      \frac{1}{\cosh X} & (X > X_0)
\end{cases}
\label{aX}	
\end{align}
As before, we use $d\Om^2$($d\om^2$) to denote the metric of the
three-sphere(two-sphere) respectively.
In these coordinates, the ``thin wall" sits at $X=X_0$, {\it i.e.,}
$\phi=\phi_-$ for $X < X_0$ and $\phi=\phi_+$ for $X>X_0$.
The radius of the $S^4$ ``outside" the bubble is set to $1$.

\begin{figure}[!b]
\begin{minipage}[b]{0.5\linewidth}
\centering \includegraphics[width=6cm]{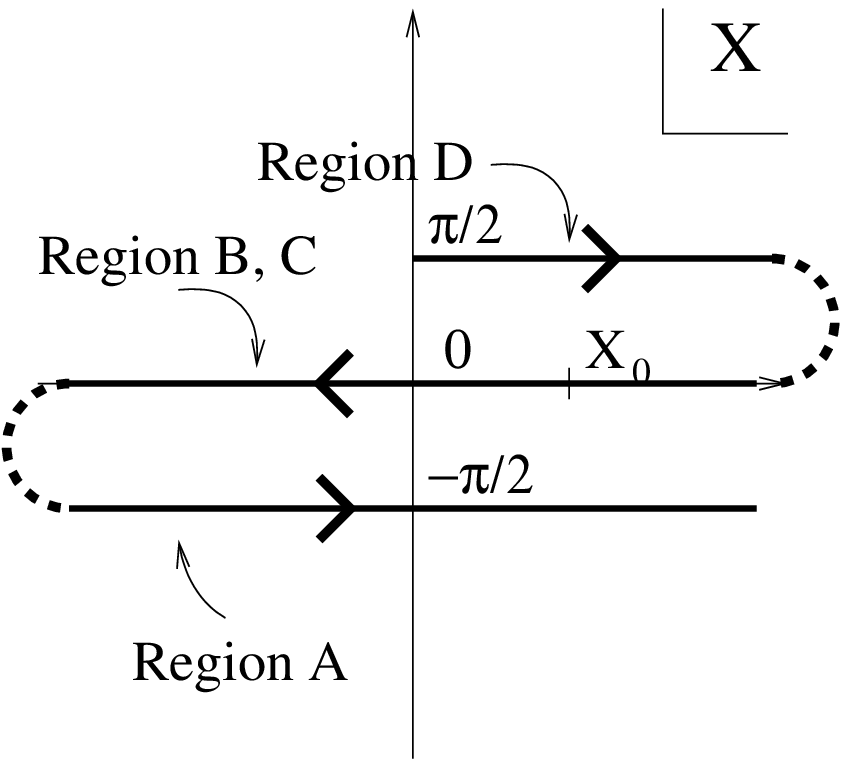}
\caption{\small The X contour}\label{fig:X}
\end{minipage}%
\begin{minipage}[b]{0.5\linewidth}
\centering \includegraphics[width=6cm]{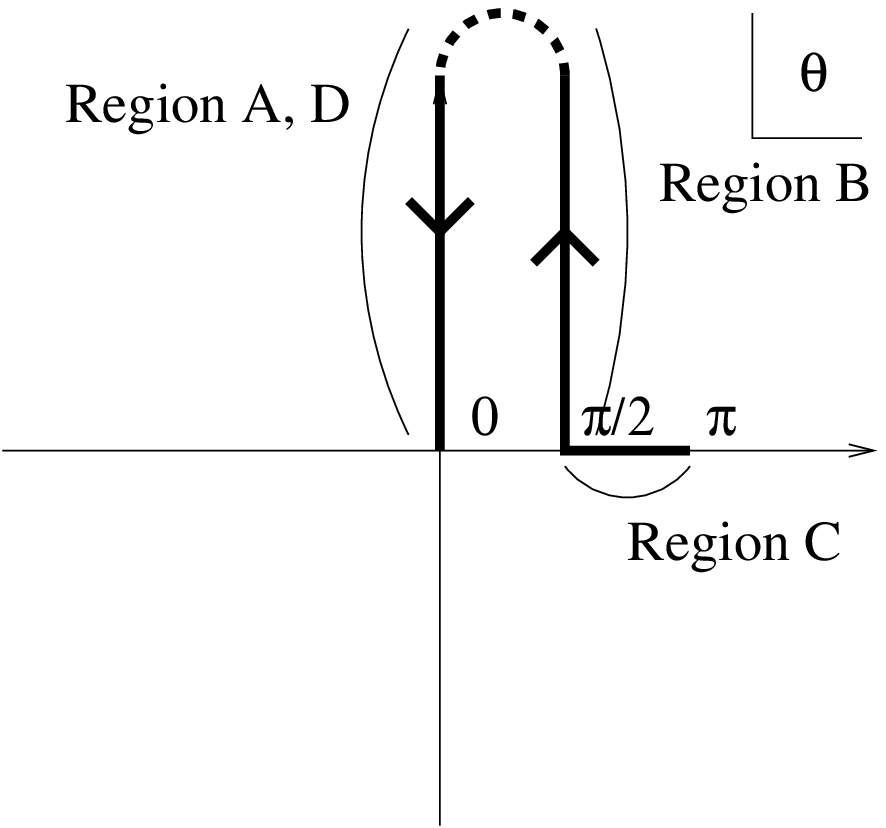}
\caption{\small The $\theta$ contour}\label{fig:th}
\end{minipage}
\end{figure}

The coordinate X runs over the contour
\be
\begin{cases}
{\rm Im}~X = i \pi/2,~ {\rm Re}~ X \geq 0 \\
{\rm Im}~ X=0 \\ 
{\rm Im}~ X = -i \pi/2 \,,
\end{cases}
\ee
and $\theta$ is defined for the contour
\be
\begin{cases}
 {\rm Im}~ \theta =0,~~~\pi/2 \leq {\rm Re}~ \theta \leq \pi \\
 {\rm Re}~ \theta =\pi/2,~~~ {\rm Im}~ \theta \geq 0 \\
 {\rm Re}~ \theta =0,~~~ {\rm Im}~ \theta \geq 0
\end{cases}
\ee
on the complex plane.

To extend the definition of $a(X)$ over this contour
we define
\begin{align}
 a(X) = \frac{e^{X-X_0 }}{\cosh X_0 }
\end{align}
on the contour ${\rm Im}~ X = -i \pi/2$, and
\begin{align}
 a(X) = \frac{1}{\cosh X}
\end{align}
on the contour ${\rm Im}~ X = i \pi/2,~ {\rm Re}~ X \geq 0$.

The analytic continuation required to obtain
the flat FRW region(let us call this region, region A) is
\begin{align}
	X = T - i \pi /2,~~~~ \theta = i R ~(R \geq 0) \,,
\label{anconA}
\end{align}
which sends slices of three-spheres to slices of
three-hyperbolic spaces.
This yields the metric
\begin{align}
\begin{split}
 ds^2 &=  (\frac{e^{T-X_0 }}{\cosh X_0 })^2 ( - dT^2 + dR^2 + \sinh^2 R d \Omega^2 ) \\
 &= (\frac{e^{T-X_0 }}{\cosh X_0 })^2 ( - dT^2 + d\HH^2 ) \,,
\end{split}
\end{align}
which provides the metric for the FRW region inside the bubble.
As before, $d\HH^2$ denotes the metric for the
three-dimensional hyperbolic space.

The space-like region(region B) of the CDL background is given by
\begin{align}
	\theta \rightarrow i t + \pi/2~(t \geq 0) \,,
\label{anconB}
\end{align}
which results in the metric
\begin{align}
	ds^2 &= a^2 (X) (dX^2 - d t^2 + \cosh^2 t d \om^2 ) \,.
\end{align}

Let us denote the Euclidean manifold patched to the space-like region by
region C. Its metric is given by
\begin{align}
 ds^2 &= a^2 (X) (dX^2 + d \theta^2 + \sin^2 \theta d \om^2 ) \,.
\end{align}

\begin{figure}[!t]
\centering \includegraphics[width=8cm]{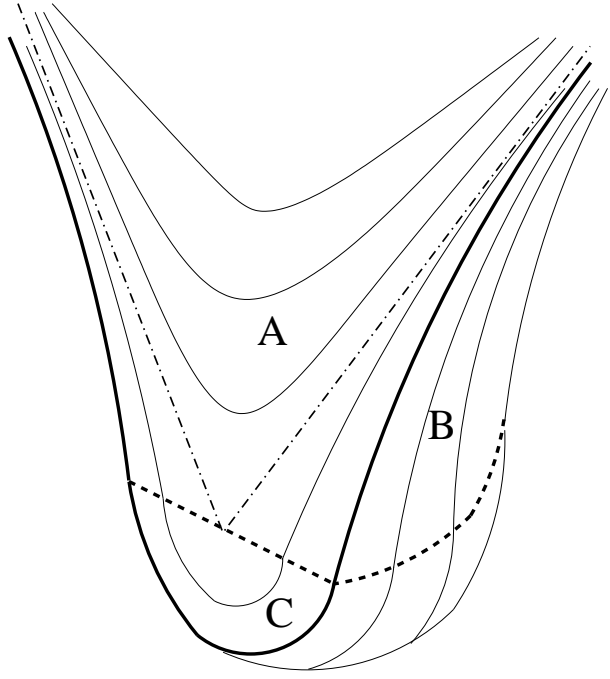}
\caption{\small A two-dimensional thin-wall CDL instanton.
Region A and B are divided by the dash/dotted line while
region B and C are divided by the dashed line.
The bold line is the domain wall $X=X_0$.
The contours are equal $X$ lines.}\label{fig:d2}
\end{figure}

Finally, we denote the de Sitter FRW region patched to the
space-like region, D. It is obtained by the analytic continuation
\begin{align}
	X = \tau + i \pi /2,~~~~ \theta = i \rho ~(R \geq 0) \,,
\label{anconD}
\end{align}
which sends slices of three-spheres to slices of three-hyperbolic spaces.
This yields the metric
\begin{align}
\begin{split}
 ds^2 &=  \frac{1}{\sinh^2 \tau } ( - d\tau^2 + d\rho^2 + \sinh^2 \rho d \om^2 ) \\
 &= \frac{1}{\sinh^2 \tau } ( - d\tau^2 + d\HH^2 ) \,,
\end{split}
\end{align}
which provides the metric for the de Sitter FRW region.

To summarize, the analytically continued coordinates of the four regions
embedded in five-dimensional ``space" are
\begin{align}
\begin{cases}
 &({\sinh X_0 \ov \cosh X_0},-i\frac{e^{T-X_0 }}{\cosh X_0 }\cosh R,
 \left( \frac{e^{T-X_0 }}{\cosh X_0 } \sinh R \right)  \om) \quad \text{A} \\
 &({\sinh X_0 \ov \cosh X_0},-i\frac{e^{X-X_0 }}{\cosh X_0 }\sinh t,
 \left(\frac{e^{X-X_0 }}{\cosh X_0 } \cosh t \right) \om) \quad \text{B, }X<X_0 \\
 &({\sinh X_0 \ov \cosh X_0},\frac{e^{X-X_0 }}{\cosh X_0 } \cos \theta,
 \left(\frac{e^{X-X_0 }}{\cosh X_0 } \sin \theta \right) \om) \quad \text{C, }X<X_0 \\
 &({\sinh X \ov \cosh X}, -i{1 \ov \cosh X} \sinh t,
 \left({1 \ov \cosh X} \cosh t \right) \om) \quad \text{B, }X>X_0 \\
 &({\cosh \tau \ov \sinh \tau}, {1 \ov \cosh X} \cos \theta,
 \left({1 \ov \cosh X} \sin \theta \right) \om) \quad \text{C, }X>X_0 \\
 &({\cosh \tau \ov \sinh \tau}, -i{1 \ov \sinh \tau} \cosh \rho,
 \left({1 \ov \sinh \tau} \sinh \rho\right)  \om) \quad \text{D }
\end{cases}
\end{align}
where $\om$ denotes the embedded coordinates of the two-sphere.
Imaginary coordinates have been used to denote the time-like direction.
The scalar field $\phi$ takes the value $\phi_-$ in region
A and parts of regions B, C with $X<X_0$.
It takes the value $\phi_+$ in region D
and parts of regions B, C with $X>X_0$.

\begin{figure}[!t]
\leavevmode
\begin{center}
\epsfysize=8cm  % this number is adjustable , for the size of the image
\epsfbox{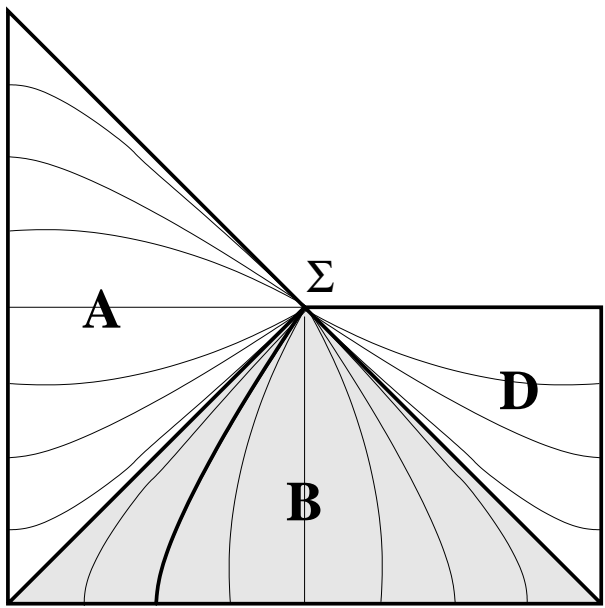}
\end{center}
\caption{\small The Penrose diagram for the Lorentzian
signature regions of the thin-wall Coleman-De Luccia instanton.
The bold curve in the grey region is the bubble wall. The space is flat
to the left of the wall, and deSitter to the right side of the wall.
}
\label{BubblePD}
\end{figure}

Region A and B are patched together along the lightcone
$T=-\infty, R=\infty$ and $X=-\infty, t=\infty$ so that
\be
 T+R = X+t \,. \label{patch1}
\ee
Region B and C are patched together at $t=0$ and $\theta=\pi/2$.
Region B and D are patched together at $X=\infty, t=\infty$ and $\tau=\infty, \rho=\infty$
so that
\be
 X-t = \tau-\rho \label{patch2} \,.
\ee
Also note that the regions A, B and C share the common point,
$(-{\sinh X_0 \ov \cosh X_0},0,0)$ where for region A, is at $T=-\infty$
for finite $R$, and for region B, C is at $X=-\infty$ and $t, \theta$ are finite.
Similarly, the regions B, C and D share the common point,
$(1,0,0)$ where for region A, is at $\tau=\infty$ for finite $\rho$,
and for region B, C is at $X=\infty$ for finite $t$ and $\theta$.
This can be summarized by figures \ref{fig:X} and \ref{fig:th}.

As noted in section \ref{s:objective},
the analytic continuation we use differs from the standard conventions
used in the literature \cite{FSSY,GrattonTurok,Park}.
We have found this different convention convenient to use for
describing individual modes in this background,
as we found it easy to define the patching \eq{patch1} and \eq{patch2}
of the various regions using this choice.

The diagram for a two-dimensional thin-wall CDL instanton
embedded in three-dimensional space is given in figure \ref{fig:d2}.
The contours are equal $X$ lines.
By obvious dimensional limitations, region D is not depicted.

We can also draw a Penrose diagram of this space.
Figure \ref{BubblePD} is a Penrose diagram of the
regions with Lorentzian signature.
The thin curves inside this region denotes
constant $T$ slices which are $\HH^3$'s.
The asymptotic boundary for region A is at
space-like infinity, $R \rightarrow \infty$.
We have denoted this boundary $\Sigma$
in the introduction.

\subsection{Eigenmodes of the Radial Potential} \label{ss:modes}

In this section, we compute the eigenmodes of the radial potential
for the scalar field.
To do so, we first compute the potential $U(X)$ for the thin-wall
metric $a(X)$. It is given by
\begin{align}
U(X) &= \begin{cases}
1   & (X < X_0) \\
1-{2-m^2 \ov \cosh^2 X}  & (X > X_0)
\end{cases}
\end{align}

We know two sets of unbounded solutions
for the equation
\be
 (-\p_X^2 + U(X)) \Psi_k (X) = (1 +k^2 )\Psi_k (X)
\label{radial}
\ee
for the thin-wall case which are
\begin{align}
\Psi_k &= \begin{cases}
e^{ikX} + \sR(k) e^{-ikX}   & (X < X_0) \\
\sT(k) \phi_k (X)  & (X > X_0)
\end{cases}
\\[10pt]
\Phi_k &= \begin{cases}
\sT(k) e^{ikX}   & (X < X_0) \\
\phi_{-k} (X) - {\sR(-k) \sT(k) \ov \sT(-k)} \phi_k (X)   & (X > X_0)
\end{cases}
\end{align}
$\phi_k$ is defined to be
\begin{align}
 \phi_k (X) &= e^{ikX} F(-\nu,\nu+1;1-ik;{1-\tanh X \ov 2})
\end{align}
$F$ is the hypergeometric function ${}_2F_1$.
Note that for $X \rightarrow \infty$,  $\phi_k (X) \rightarrow e^{ikX}$.
$\nu$ is defined to be
\be
\nu \equiv (1-\epsilon) \equiv {\sqrt{9-4m^2}-1 \ov 2} \approx 1-{m^2 \ov 3} \,.
\ee

The boundary conditions we must solve to obtain $\sR$ and
$\sT$ are
\begin{align}
e^{ikX_0}+\sR(k) e^{-ikX_0} &= \sT(k) \phi_k (X_0) \\
ik e^{ikX_0}-ik \sR(k) e^{-ikX_0} &= \sT(k) \left[ \phi_k' (X_0)
+(\tanh X_0+1)\phi_k (X_0) \right] \,.
\end{align}
The last term of the last equation comes from being careful
with the singularity of
$a''(X)/a(X)$ at $X=X_0$.

The reflection and transmission coefficients are given by
\begin{align}
 \sR(k) &= e^{2ikX_0}
 \left[ {(t-1){ b_k (t) + \epsilon (t-1)(d_k (t)-b_k (t) ) } \ov
 (ik+1) c_k (t) -\epsilon \left[ t c_k(t)+{t(t-1) \ov 1-\epsilon}c_k'(t) \right] } \right] \label{r1}\\
 \sT(k) &= {ik \ov (ik+1) c_k (t) -\epsilon \left[ t c_k(t)+{t(t-1) \ov 1-\epsilon}c_k'(t) \right]} \label{t1}
\end{align}
We have defined
\begin{align}
b_k (x) &= F(-\nu+1,\nu+1;1-ik;x)  \\
c_k (x) &= F(-\nu,\nu;1-ik;x)\\
c'_k (x) &= {d \ov dy}F(-\nu,\nu;1-ik;y) |_{y=x}\\
d_k (x) &= F(-\nu,\nu+1;1-ik;x)
\end{align}
$t$ is defined as
\be
t \equiv {L \ov 2} \equiv {e^{-X_0} \ov 2 \cosh X_0} \,.
\ee
It lies in the range
\be
0<t<1\,.
\ee

In the $m \rightarrow 0$ limit, we find that
\begin{align}
 \phi_k (X) &\ra e^{ikX} \left( {\tanh X-ik \ov 1-ik} \right) \\
 \sR(k) &\ra e^{2ikX_0}{(-1+ik)(1+\tanh X_0 ) \ov (1+ik)(\tanh X_0 +1-2ik)}\\
 \sT(k) &\ra {2ik(1-ik) \ov (1+ik)(\tanh X_0 +1 -2ik)} \,.
\end{align}

Under the assumptions we have made, the analytic continuation
of the modes to the interior of the bubble(region A) is straightforward.
$\Phi_k(X)/\sT(k)$ can be analytically continued to
\[
\fbox{
\addtolength{\linewidth}{-2\fboxsep}%
\addtolength{\linewidth}{-2\fboxrule}%
\begin{minipage}{\linewidth}
\be
{\Phi_k(X) \ov \sT(k)} \rightarrow e^{-ik(T-i\pi/2)}
\ee
\end{minipage}\nonumber
}
\]
under $X \rightarrow T-i\pi/2$.

Plugging this into the holographic expansion \eq{holexp},
and using the fact that $\ta(T) = Le^{T}$,
we indeed arrive at an holographic expansion of the form \eq{holform}.
Recall that
\begin{align}
\begin{split}
& \langle \vh(T_1,\HH_1)\vh(T_2,\HH_2)\vh(T_3,\HH_3) \rangle \\
&= \sum_\sof{\s_i} \sum_\sof{n_{r_i}^{\s_i}} 
F^\sof{\s_i}_\sof{n^{\s_i}_{r_i}+1} (\sof{T_i}) ( \prod_i  z^{n^{\s_i}_{r_i}+1} )
 \tau_\sof{n^{\s_i}_{r_i}+1} (\sof{\vec{x}_i})
+ \text{(log terms)} \,.
\label{premassless}
\end{split}
\end{align}
for
\begin{align}
\begin{split}
F^\sof{\s_i}_\sof{n^{\s_i}_{r_i}+1}(\sof{T_i})
&=\Res{\sof{k_i}=\sof{i\s_i n^{\s_i}_{r_i}}}
\left( \prod_i {\Phi_{k_i} (T_i-i\pi/2)\ov \ta(T_i) \sT(k_i) } \right) \hS_\sof{\s_i} (\sof{k_i}) \\
&=\left( {1 \ov L}\right)^3  \left( \prod_i e^{[-1+\s_i(\Delta^{\s_i}_{r_i}-1)]T_i} \right) \Res{\sof{k_i}=\sof{i\s_i n^{\s_i}_{r_i}}}
\hS_\sof{\s_i} (\sof{k_i}) \,.
\end{split}
\end{align}
where we have used $\Delta^{\s}_{r} = n^{\s}_{r}+1$ as before.
Therefore it is clear that \eq{premassless} can be written in the form
\eq{holform}:
\[
\fbox{
\addtolength{\linewidth}{-2\fboxsep}%
\addtolength{\linewidth}{-2\fboxrule}%
\begin{minipage}{\linewidth}
\begin{align}
\begin{split}
& \langle \vh(T_1,\HH_1)\vh(T_2,\HH_2)\vh(T_3,\HH_3) \rangle \\
&= \sum_\sof{\s_i} \sum_\sof{n_{r_i}^{\s_i}} C^\sof{\s_i}_\sof{n_{r_i}^{\s_i}}
 \left( \prod_i e^{[-1+\s_i(\Delta^{\s_i}_{r_i}-1)]T_i} z^{n^{\s_i}_{r_i}+1} \right)
 \tau_\sof{n^{\s_i}_{r_i}+1} (\sof{\vec{x}_i})
+ \text{(log terms)} \,.
\label{massless}
\end{split}
\end{align}
\vskip2pt
\end{minipage}\nonumber
}
\]
The structure coefficients can be identified with residues
of $\hS_\sof{\s_i} (\sof{k_i})$, {\it i.e.,}
\[
\fbox{
\addtolength{\linewidth}{-2\fboxsep}%
\addtolength{\linewidth}{-2\fboxrule}%
\begin{minipage}{\linewidth}
\be
C^\sof{\s_i}_\sof{ \Delta^{\s_i}_{r_i}}=
-\left( {\prod_i e^{-i\pi \Delta^{\s_i}_{r_i}} \ov L^3} \right)
\Res{\sof{k_i}=\sof{i\s_i (\Delta^{\s_i}_{r_i}-1)}}
{\cosh(\pi \sum_i \s_i k_i/2) c_\sof{ 1- i \s_i k_i} \ov 64 \pi^2 \prod_i \sinh k_i \pi}
S(\sof{k_i}) \,.
\ee
\vskip2pt
\end{minipage}\nonumber
}
\]
The structure coefficient of a term coming from a generic pole at
$\sof{i\s_i (\Delta_i-1)} \in H^{\s_1}_{k_1} \times H^{\s_2}_{k_2} \times H^{s_3}_{k_3}$
can be related to the wavefunction overlap in an even simpler way:
\[
\fbox{
\addtolength{\linewidth}{-2\fboxsep}%
\addtolength{\linewidth}{-2\fboxrule}%
\begin{minipage}{\linewidth}
\be
C^\sof{\s_i}_\sof{ \Delta_i}=
-{ i \ov 64 \pi^5 L^3}  e^{-i\pi \sum_i \Delta_i/2}
\sin({\pi \sum_i \s_i \Delta_i \ov 2})
S(\sof{i\s_i (\Delta^{\s_i}_{r_i}-1)}) c_\sof{ \Delta_i }  \,.
\ee
\vskip2pt
\end{minipage}\nonumber
}
\]

\subsection{The Wavefunction Overlap} \label{ss:overlap}

We compute the wavefunction overlap
\be
S(\sof{k_i}) =  \int_{-\infty}^\infty dX a(X) \Psi_{k_1}(X)\Psi_{k_2}(X)\Psi_{k_3}(X) \,,
\ee
for the thin-wall instanton in this section.

As we assumed that
\begin{align}
{d^3 \mathcal{V} \ov d\vh^3}|_{\vh=0} &=0 \quad \text{at}~\phi=\phi_-  \\
{d^3 \mathcal{V} \ov d\vh^3}|_{\vh=0} &=\lambda \quad \text{at}~\phi=\phi_+ \,,
\end{align}
the overlap is given by
\begin{align}
\begin{split}
S(\sof{k_i}) &=  \lambda \int_{X_0}^\infty dX a(X) \Psi_{k_1}(X)\Psi_{k_2}(X)\Psi_{k_3}(X) \\
&=\lambda  \sT(k_1) \sT(k_2)  \sT(k_3) \int_{X_0}^\infty {dX \ov \cosh X}
\phi_{k_1} (X) \phi_{k_2} (X) \phi_{k_3} (X) \,.
\end{split}
\end{align}

We have been able to find a series expansion for
the integrand. We note that
\begin{align}
\begin{split}
\phi_k (X) &= {e^{ikX}} F(-\nu,\nu+1;1-ik;{e^{-2X} \ov 1+e^{-2X}}) \\
&= {e^{ikX}} \sum_{m=0}^\infty {(-\nu)_m (\nu+1)_m \ov m! (1-ik)_m}
\left( {e^{-2X} \ov 1+e^{-2X}} \right)^{m} \,,
\end{split}
\end{align}
where, as before,
\be
(a)_n \equiv a (a+1) \cdots (a+n-1)\,.
\ee
Using the fact that
\begin{align}
\begin{split}
\FF_{N,K} (X_0)
&\equiv \int_{X_0}^\infty {{dX} \ov \cosh X} e^{iKX} {e^{-2NX} \ov (1+e^{-2X})^N}\\
&=e^{(iK-2N-1)X_0} \sum_{m=0}^\infty {(N+1)_m \ov m! (-{iK\ov 2}+{1 \ov 2}+N+m) } (-e^{-2X_0})^m \\
&={e^{(iK-2N-1)X_0} \ov -{iK\ov 2}+N+{1 \ov 2}}
F(N+1,-{iK\ov 2}+{1 \ov 2}+N; -{iK\ov 2}+{3 \ov 2}+N;-e^{-2X_0}) \,,
\end{split}
\end{align}
$S(\sof{k_i})$ can be shown to be
\begin{align}
\begin{split}
S (\sof{k_i})
= \lambda
\sum_\sof{m_i}
\left[ \prod_i  {(-\nu)_{m_i} (\nu+1)_{m_i} \ov m_i ! (1-ik_i)_{m_i}} {\sT(k_i) } \right]
\FF_{(m_1+m_2+m_3),K}(X_0)
\,.
\end{split}
\label{sleading}
\end{align}
The sum of $m_i$ runs over the non-negative integers.
We have defined
\be
K \equiv k_1 + k_2 + k_3 \,.
\ee

We point out two non-trivial cancellations
related to $S(\sof{k_i})$. The zeros of $\sT(k_i)$ at $k_i = -in_i$
are cancelled by the zeros of $(1-ik_i)_m$ for $m \geq n_i$.
Meanwhile,
\be
\Res{K=-i(2n+1)} \FF_{N,K} (X_0)
=\begin{cases}
 0   & (n < N) \\
2i(-1)^{n-N} {n! \ov N! (n-N)!}  & (n \geq N)
\end{cases}
\ee
This cancels the zero of $\cos(\pi\sum \s_i k_i/2)$
and makes the three-point functions of dimension
$\sof{\Delta_i}$ with odd $\sum_i (\Delta_i-1)$
contribute in the holographic expansion.

\section{The Massive Scalar in the Thin-wall Limit} \label{s:m1thinwall}

In this section, we compute the three-point function of a massive scalar
in the thin-wall CDL instanton background introduced in section \ref{ss:thinwall}.
We assume the potential for the scalar $\vh$ has the expansion
\be
\mathcal{V}(\vh) =
{1 \ov 2} m^2 \vh^2 + {1 \ov 6} \lambda \vh^3 + \OO(\vh^4) \,.
\ee
Such a potential can be obtained by a potential independent
of other background fields such as \eq{pot2}.
The radial potential $U(X)$ derived from this potential
is regular for small enough $m$, as shown in appendix \ref{aps:m1},
and hence we can use the results of the previous sections.

As in the previous section we compute the
\begin{enumerate}
\item The eigenmodes $\Phi_k, \Psi_k$ of the radial potential.
\item The wavefunction overlap $S(\sof{k_i})$ in the radial direction.
\end{enumerate}
for this scalar.
We first present the eigenmodes $\Phi_k, \Psi_k$
and their analytic continuation in section \ref{ss:m1modes}.
Here we identify the poles of ${\Phi_k /\sT(k)}$ and discuss
their effects on the holographic expansion.
We also discuss the analytic structure of $\Phi_k, \Psi_k$
and the reflection/transmission coefficient along the way.
We compute $S(\sof{k_i})$ in section \ref{ss:m1overlap}.

\subsection{Eigenmodes of the Radial Potential} \label{ss:m1modes}

In this section, we compute the eigenmodes of the radial potential
for the scalar field.
To do so, we first compute the potential $U(X)$ for the thin-wall
metric $a(X)$. This is given by
\begin{align}
U(X) &= \begin{cases}
1+\mu^2 e^{2X}   & (X < X_0) \\
1-{2-m^2 \ov \cosh^2 X}  & (X > X_0)
\end{cases}
\end{align}
where 
\be
\mu =mL, \qquad L= {e^{-X_0} \ov \cosh X_0}\,.
\ee

We know two sets of unbounded solutions
for the equation
\be
 (-\p_X^2 + U(X)) \Psi_k (X) = (1 +k^2 )\Psi_k (X)
\ee
for the thin-wall case. They are given by
\begin{align}
\Psi_k &= \begin{cases}
\psi_k (X) + \sR(k) \psi_{-k}(X)   & (X < X_0) \\
\sT(k) \phi_k (X)  & (X > X_0)
\end{cases}
\\[10pt]
\Phi_k &= \begin{cases}
\sT(k) \psi_{-k}(X)   & (X < X_0) \\
\phi_{-k} (X) - {\sR(-k) \sT(k) \ov \sT(-k)} \phi_k (X)   & (X > X_0)
\end{cases}
\end{align}
$\psi_k, \phi_k$ are defined to be
\begin{align}
 \psi_k (X) &= ({\mu \ov 2})^{-ik} \Ga(ik+1) I_{ik} (\mu e^{X})\\
 \phi_k (X) &= e^{ikX} F(-\nu,\nu+1;1-ik;{1-\tanh X \ov 2})
\end{align}
$F$ is the hypergeometric function ${}_2F_1$
and $I$ is the modified Bessel function.
A before, $\nu$ and $\epsilon$ are defined to be
\be
\nu\equiv 1-\epsilon
 \equiv {\sqrt{9-4m^2}-1 \ov 2} \approx 1-{m^2 \ov 3} \,.
\ee

The reflection and transmission coefficients are given by
\begin{align}
 \sR(k) &= \left[ {\psi_k (X_0) \ov \psi_{-k} (X_0)} \right]
 \left[ {(t-1){ b_k (t) + \epsilon \{ (t-1)(d_k (t)-b_k (t) ) \} +\mu {e^{X_0} d_k (t) g_k(X_0) \ov 2\psi_k(X_0)}} \ov
 (ik+1) c_k (t) -\epsilon \{ t c_k(t)+{t(t-1)  \ov (1-\epsilon)}c_k'(t) \}- \mu {e^{X_0} d_k (t) g_{-k}(X_0) \ov 2\psi_{-k}(X_0)} } \right] \\
 \sT(k) &= {ik e^{-ikX_0}\ov \psi_{-k}(X_0)  \left[ (ik+1) c_k (t) -\epsilon \{ t c_k(t)+{t(t-1)  \ov (1-\epsilon)}c_k'(t) \}
 - \mu {e^{X_0} d_k (t) g_{-k}(X_0) \ov 2\psi_{-k}(X_0)}\right] } 
\end{align}
We have defined $b,c,c'$ and $d$ as before and
\begin{align}
g_k (X) &=  ({\mu \ov 2})^{-ik} \Ga(ik+1) {I_{ik+1} (\mu e^{X})} \,.
\end{align}
As before, $t$ is defined as
\be
t \equiv {L \ov 2} \equiv {e^{-X_0} \ov 2 \cosh X_0}\,,
\ee
and is in the range $0<t<1$.

Under the assumptions we have made, the analytic continuation
of the modes to the flat FRW region (region A) is straightforward.
$\Phi_k (X)/\sT(k)$ can be analytically continued to
\[
\fbox{
\addtolength{\linewidth}{-2\fboxsep}%
\addtolength{\linewidth}{-2\fboxrule}%
\begin{minipage}{\linewidth}
\be
{\Phi_k(X) \ov \sT(k)} \rightarrow  ({i\mu \ov 2})^{ik} \Ga(-ik+1) J_{-ik} (\mu e^{T})
\label{massivePhi}
\ee
\end{minipage}\nonumber
}
\]
under $X \rightarrow T-i\pi/2$, where
$J$ is the Bessel function.

It is clear that $\Phi_k(X) / \sT(k)$ has an infinite number of poles
due to the $\Ga$ function.
They are situated at
\be
k=-in, \quad \text{$n$ is a positive integer} \,,
\ee
which are all in the lower-half plane.
When $k$ is not near these values,
\be
{\Phi_k(T-i\pi/2) \ov \sT(k)} = e^{-k\pi/2} e^{-ikT} (1+ \OO(e^T)) \,.
\ee
However, near $k=-in$ we find that
\begin{align}
\begin{split}
{\Phi_k(T-i\pi/2) \ov \sT(k)} &= {i \ov k+in} \left( {\mu^{2n} e^{in\pi/2} \ov 4^n n! (n-1)!}\right) e^{nT} (1+\OO(e^T)) \\
&+ e^{in\pi/2} e^{-nT} (1+ \OO(e^T)) \,,
\end{split}
\end{align}
and hence
\be
\Res{k=-in} {\Phi_k(T-i\pi/2) \ov \sT(k)} \propto e^{nT}\,.
\ee

As discussed in section \ref{ss:holexp}, the contribution of
these poles are negligible in the early-time limit.
They, however, do contribute in the late time limit.
In fact, it can be shown that
\be
\Res{k=-in}{\Phi_k(T-i\pi/2) \ov \sT(k)} \propto J_n (\mu e^T)
\propto {\Phi_{in}(T-i\pi/2) \ov \sT(in)} \,,
\ee
which is exactly how the of poles in $H^{+1}_k$ scale.
Therefore, to obtain the correct late-time behavior of the holographic
expansion, one must consider contributions from
poles in all products of half-planes,
$H^{\s_1}_{k_1} \times H^{\s_2}_{k_2} \times H^{\s_3}_{k_3}$.

\subsection{The Wavefunction Overlap} \label{ss:m1overlap}

We compute the wavefunction overlap
\be
S(\sof{k_i}) =  \int_{-\infty}^\infty dX a(X) \Psi_{k_1}(X)\Psi_{k_2}(X)\Psi_{k_3}(X) \,,
\ee
for the thin-wall instanton in this section.

As we have assumed that
\begin{align}
{d^3 \mathcal{V} \ov d\vh^3}|_{\vh=0} =
\lambda \,,
\end{align}
the overlap turns out to be
\begin{align}
\begin{split}
S(\sof{k_i})&=  \lambda \int_{-\infty}^\infty dX a(X) \Psi_{k_1}(X)\Psi_{k_2}(X)\Psi_{k_3}(X) \,.
\end{split}
\end{align}
We can split $S$ as the sum of two pieces:
\begin{align}
S(\sof{k_i})&=S_1(\sof{k_i})+S_2(\sof{k_i}) \\
\begin{split}
S_1(\sof{k_i})
&= \lambda \int_{-\infty}^{X_0} dX a(X) \Psi_{k_1}(X)\Psi_{k_2}(X)\Psi_{k_3}(X)\\
&= \lambda L \int_{-\infty}^{X_0} dX e^X \prod_i (\psi_{k_i} + \sR(k_i) \psi_{-k_i})
\end{split} \\
\begin{split}
S_2(\sof{k_i})
&= \lambda \int_{X_0}^{\infty} dX a(X) \Psi_{k_1}(X)\Psi_{k_2}(X)\Psi_{k_3}(X) \\
&= \lambda \int_{X_0}^{\infty} dX {1 \ov \cosh X} \prod_i \left( \sT(k_i) \Phi_{k_i}(X) \right)
\end{split}
\end{align}

$S_2$ has been calculated in section \ref{ss:overlap}---it is
given by \eq{sleading}.
$S_1$ can be obtained straightforwardly
once
\begin{align}
\begin{split}
\int_{-\infty}^{X_0} dX e^X \psi_{k_1}(X)\psi_{k_2}(X)\psi_{k_3}(X)
\end{split}
\end{align}
is known. Using
\be
 \psi_{k}(X) =
({\mu \ov 2})^{-ik} \Ga(ik+1) I_{ik} (\mu e^{X}) = e^{ikX} \sum_{m=0}^\infty
{1 \ov m! (ik+1)_m} \left({1 \ov 4} \mu^2 e^{2X}\right)^m
\ee
one finds that
\begin{align}
\begin{split}
&\int_{-\infty}^{X_0} dX e^X \psi_{k_1}(X)\psi_{k_2}(X)\psi_{k_3}(X) \\
&=\sum_\sof{m_i}  \left[ \prod_i {\left( {\mu^2/ 4}\right)^{m_i} \ov m_i! (ik_i+1)_{m_i}} \right] 
 {e^{(iK+2(m_1+m_2+m_3)+1)X_0} \ov iK+2(m_1+m_2+m_3)+1}  \,.
\end{split}
\end{align}
The sum of $m_i$ runs over non-negative integers.
We have defined
\be
K \equiv k_1 + k_2 + k_3 \,.
\ee

\section{Summary and Discussion} \label{s:conclusion}

\subsection{Summary of Results} \label{ss:summary}

We have computed the three-point function of a
scalar in a CDL background in four dimensions
in the setup explained in section \ref{s:objective}.
When the three points are in the FRW region
with metric
\be
ds^2 = \ta(T)^2 (-dT^2 + d\HH^2)\,,
\ee
it is given by the integral:
\[
\fbox{
\addtolength{\linewidth}{-2\fboxsep}%
\addtolength{\linewidth}{-2\fboxrule}%
\begin{minipage}{\linewidth}
\begin{align}
\begin{split}
& \langle \vh(T_1,\HH_1)\vh(T_2,\HH_2)\vh(T_3,\HH_3) \rangle \\
&=
\prod_i \left( \int_C {dk_i\ov 4\pi i \sinh k_i \pi} {\Phi_{k_i} (T_i-i\pi/2) \ov \ta(T_i) \sT(k_i)} \right)
S(\sof{k_i}) \sum_{\sof{\s_i}} \s_1 \s_2 \s_3
(1+ e^{-(\sum_i \s_i k_i)  \pi})  \UU_\sof{\s_i k_i} (\sof{\HH_i}) 
\end{split}
\label{full3pt}
\end{align}
\end{minipage}\nonumber
}
\]
We have assumed that the radial potential
\be
U(X) = {a'' (X) \ov a(X) }
+a^2 (X) {\delta^2 \mathcal{V} \ov \delta \vh^2}|_{\phi=\phi_0, \vh=0}\,,
\ee
is ``regular." We have defined regularity in section \ref{ss:euc2}.
We note that $U(X)$ for both the massless and massive
scalar in a thin-wall CDL instanton background is regular.

For convenience, we have used $\sof{x_i}$ to denote the triplets
$(x_1,x_2,x_3)$ for any variable $x$.
The sum over $\s_i$ runs over the values $(+1)$ and $(-1)$.
$\UU_\sof{k_i}$ are the three-point functions on three-hyperbolic space
$\HH^3$:
\be
\UU_\sof{k_i} (\sof{\HH_i})
= \int d\HH \prod_i {e^{ik_i \ell(\HH_i,\HH)} \ov \sinh \ell(\HH_i,\HH)}\,.
\ee
$\ell$ denotes the angular distance between two points in $\HH^3$.
The contour of integration $C$ is depicted in figure \ref{f:contourp}.
It is a contour that runs along the real line with a jump
over the upmost pole of $\sT(k)$, which we soon define.

All the non-trivial data of the background and interactions
are encoded in $\Phi_{k}/\sT(k)$ and the wavefunction
overlap $S(\sof{k_i})$.
$\Phi_k$ and $\Psi_k$ are eigenmodes of the radial
potential $U(X)$
with the following asymptotic behavior:
\begin{align}
\Psi_k &\rightarrow \begin{cases}
e^{ikX} + \sR(k) e^{-ikX}   & (X \rightarrow -\infty) \\
\sT(k) e^{ikX}  & (X \rightarrow \infty)
\end{cases} \\
\Phi_k &\rightarrow \begin{cases}
\sT(k) e^{-ikX}   & (X \rightarrow -\infty) \\
e^{-ikX} - {\sR(-k) \sT(k) \ov \sT(-k)} e^{ikX}   & (X \rightarrow \infty)
\end{cases}
\end{align}
The wavefunction overlap is defined to be
\be
S(\sof{k_i}) = \int_{-\infty}^\infty dX W(X)
\Psi_{k_1}(X) \Psi_{k_2}(X) \Psi_{k_3}(X) \,,
\ee
where
\be
W(X) =
a(X) {\delta^3 \mathcal{V} \ov \delta \vh^3}|_{\phi=\phi_0, \vh=0} \,.
\ee
The functions $\Phi_{k}(T-i\pi/2)/\sT(k)$ in \eq{full3pt}
are the radial wavefunctions analytically continued to
the FRW region of the CDL instanton.

We have taken various limits of the three-point function.
In particular, if we take the points of the three-point function
near the boundary of the hyperbolic slices,
the three-point function has a holographic expansion of the form
\begin{align}
\begin{split}
&\sum_\sof{\Delta_i}
F_{\Delta_1, \Delta_2, \Delta_3} (T_1,T_2,T_3)
\UU_{\Delta_1, \Delta_2, \Delta_3} (\HH_1,\HH_2,\HH_3) \\
&+ \text{(logarithmic terms)}\,.
\end{split}
\label{holexpform}
\end{align}
$\UU_{\Delta_1, \Delta_2, \Delta_3} (\HH_1,\HH_2,\HH_3)$
is the three point function in hyperbolic space:
\be
\UU_\sof{\Delta_i} (\sof{\HH_i})
= \int d\HH \prod_i {e^{-(\Delta_i-1) \ell(\HH_i,\HH)} \ov \sinh \ell(\HH_i,\HH)}\,.
\ee
The logarithmic terms have factors of $T$ and $\ln \ell$ multiplied
to $\UU$.
At early times, {\it i.e.,} when $T_i \rightarrow -\infty$,
\be
F_{\Delta_1, \Delta_2, \Delta_3} (T_1,T_2,T_3)
\ra C^{-1,-1,-1}_{\Delta_1, \Delta_2, \Delta_3}
e^{-\Delta_1 T_1} e^{-\Delta_1 T_2} e^{-\Delta_1 T_3}\,.
\ee

Let us denote $H^{\s_i}_{k_i}$ to be the upper-half of the
complex $k_i$ plane divided by $C$ when $\s_1 =+1$
and the lower-half when $\s_2=-1$.
Then, the coefficients
$C^{-1,-1,-1}_{\Delta_1, \Delta_2, \Delta_3}$
are proportional to the analytically continued wavefunction overlap
\begin{align}
C^{-1,-1,-1}_{\Delta_1, \Delta_2, \Delta_3}
\propto
\sin {\pi (\sum_i \Delta_i) \ov 2}S(-i(\Delta_1-1),-i(\Delta_2-1),-i (\Delta_3-1))
\end{align}
for $\sof{-i(\Delta_i-1)} \in H^{-1}_{k_1} \times H^{-1}_{k_2} \times H^{-1}_{k_3}$.

We have also expanded the three-point function on
the past lightcone of the FRW patch in spherical harmonics.
The results are given by equations
\eq{harmexp} and \eq{harmcoeff}.
%These equations can be applied for any regular potential $U(X)$
%regardless of the pole structure of $\Phi_k /\sT(k)$.
The $B(\sof{k_i,l_i,m_i})$ in the second equation---defined in
\eq{defB} as the integral of a triple product of eigenmodes
on $\HH^3$---are structure functions on $\HH^3$,
much like the Wigner coefficients on $S^2$.
All the non-trivial data of the coefficients of the harmonic
expansion is encoded in the wavefunction overlap $S(\sof{k_i})$,
as expected.

We have identified $\Phi_k$ and $\Psi_k$,
and have computed $S(\sof{k_i})$ for a massless and massive
scalar in a thin-wall background in sections \ref{s:m0thinwall}
and \ref{s:m1thinwall}, respectively.
We have assumed that the thin-wall divides the CDL instanton
into a flat and de Sitter region.

In the massless case, the analytic continuation of the
functions ${\Phi_k(X) / \sT(k)}$ to the FRW region are given by
\be
{\Phi_k(X) \ov \sT(k)} \rightarrow  e^{-k\pi/2}e^{-ikT} \,.
\ee
Hence the expansion \eq{holexpform}
can be written in the form
\begin{align}
\begin{split}
&\sum_{\sof{\Delta_i,\s_i}}
C^{\s_1,\s_2,\s_3}_{\Delta_1, \Delta_2, \Delta_3}
\left(\prod_{i=1}^3 e^{[-1+\s_i(\Delta_i-1)]T_i} \right)
\UU_{\Delta_1, \Delta_2, \Delta_3} (\HH_1,\HH_2,\HH_3) \\
&+ \text{(logarithmic terms)}\,.
\end{split}
\label{holexpcon}
\end{align}
Each term for a given $\sof{\s_i}$ label
comes from the residue
of a pole $(i\s_1 (\Delta_1-1),i\s_2 (\Delta_2-1),i\s_3 (\Delta_3-1))$
of the integrand of \eq{full3pt} in
$H^{\s_1}_{k_1} \times H^{\s_2}_{k_2} \times H^{\s_3}_{k_3}$.
The coefficients
$C^{\s_1,\s_2,\s_3}_{\Delta_1, \Delta_2, \Delta_3}$
of generic terms are proportional to the analytically
continued wavefunction overlap:
\be
C^{\s_1,\s_2,\s_3}_{\Delta_1, \Delta_2, \Delta_3}
\propto
\sin \left( {\pi \sum_i \s_i \Delta_i \ov 2} \right)  S(i\s_1 (\Delta_1-1),i\s_2 (\Delta_2-1),i\s_3 (\Delta_3-1))\,.
\ee

In the massive case, the analytic continuation of the
function ${\Phi_k(X) / \sT(k)}$ to the FRW region is given by
a Bessel function
\be
{\Phi_k(X) \ov \sT(k)} \rightarrow  ({i\mu \ov 2})^{ik} \Ga(-ik+1) J_{-ik} (\mu e^{T}) \,.
\ee
The terms of the holographic expansion 
\eq{holexpform} have exponential $T$ scaling at early times,
as expected. The early-time behavior can be determined
by the contribution of poles of the integrand in
$H^{-1}_{k_1} \times H^{-1}_{k_2} \times H^{-1}_{k_3}$.
Due to the behavior of Bessel functions for large $T$,
however, one must take all the poles into account to
understand the late-time behavior of the correlator.

\subsection{Discussion} \label{ss:discussion}

It is satisfying to see that a holographic expansion of
the three-point function exists for a scalar that is massless
on the flat side of the bubble.
If we assume that there is a field-operator correspondence
\be
\phi \rightarrow
\sum_{\Delta,\pm} e^{(-1\pm(\Delta-1)) T} \OO^\pm_{\Delta}
= \sum_{\Delta} e^{(\Delta-2) T} \OO^+_{\Delta}
+\sum_{\Delta} e^{-\Delta T} \OO^-_{\Delta} \,,
\label{fieldopdis}
\ee
the structure coefficients of three-point functions
of these operators are given essentially by the analytic continuation
of the wavefunction overlap, {\it i.e.,}
\be
C^{\s_1,\s_2,\s_3}_{\Delta_1,\Delta_2,\Delta_3}
\propto
\sin \left( {\pi \sum_i \s_i \Delta_i \ov 2} \right) S(\sof{i\s_i(\Delta_i-1)}) \,.
\ee
Although we have obtained an expression for $S(\sof{k_i})$
as a series sum \eq{sleading}, we 
have not examined its structure closely.
It would be interesting to study the structure of
these coefficients in detail given the mass $m^2$ of the scalar
on the de Sitter side of the wall to get a picture of
the nature of the operators $\OO^{\pm}$.

The fact that the massless scalar can be written in the
form \eq{holexpcon} on the flat FRW patch does not
depend on the thin-wall limit.
It depends, however, on the assumption that the FRW
patch is flat. The holographic expansion of a massless
scalar is expected to be
modified when this assumption is relaxed.
It seems, however, plausible that as long as the FRW patch
is asymptotically flat, we would be able to extract
the correspondence \eq{fieldopdis} by taking early-time
and late-time limits of individual points of the correlator.
It would be interesting to verify such expectations.

We have not said much about the late-time behavior of the massive
correlators in this paper. Unlike the case of the massless scalar,
the correlators for the massive scalar behave non-trivially
at late times. From \eq{massivePhi}, the late time behavior of correlators
can be deduced from the asymptotic behavior of Bessel functions
at large arguments. When $T \ra \infty$,
\be
J_\nu (\mu e^{T}) \ra \sqrt{2 \ov \mu\pi} e^{-T/2} \cos (\mu e^T -{\pi \nu \ov 2} -{\pi \ov 4}) \,.
\label{latetime}
\ee
Therefore one may expect that the late-time holographic expansion
of the massive correlators have oscillatory behavior dampened by
$e^{-3T_i/2}$ with respect to each $T_i$.
This may well be the case, but it must be checked.
Since the exponential scaling of terms coming from
$H^{+1}_k$ and $H^{-1}_k$ are the same at late times,
some non-trivial cancellation might
occur to give some other $T$ dependent behavior at large $T$.
Once the asymptotic behavior of the correlator at late times is
established, the task of modifying the conjectured
correspondence \eq{fieldopdis} to accommodate massive fields
can be addressed.
At the moment, there does not seem to be an obvious way
to generalize the field-operator correspondence if the
late time behavior is given by \eq{latetime}.
We leave investigation of such issues to future work.

There are some calculations in CDL models
that can be carried out as natural extensions of
the current calculation.
The most interesting ones are the three-point functions
that involve the inflaton $\phi$ and the metric.
The calculation involving the inflaton is subtle, as its
fluctuation mixes
with metric fluctuations.
Once, however, the mixing is sorted out,
three-point functions involving
the inflaton can be computed readily
by methods of the current paper.

Three-point functions involving the graviton can also
be carried out by a straightforward generalization
of the current calculation. This is because
the calculation of section \ref{ss:ancons3} of analytically
continuing a three-point function on the sphere to
a three-point function on hyperbolic space can
readily be generalized to tensor fluctuations.
Since we expect the stress-energy tensor
and the operator responsible for geometric fluctuations---the
``graviton"---of the boundary CFT to be in the tower
\eq{fieldopdis} of operators that correspond to the
metric field \cite{FSSY,FSSY2,FSSY3,FSSY4,FSSY5,Park},
these calculations will be crucial in
extracting data of the conjectured boundary CFT.

It would be particularly interesting to use our calculation
to study FRW backgrounds in string theory,
such as those constructed in \cite{KlebanRedi} or more recently
in \cite{DHMST}.
There are also some interesting analytic CDL solutions
constructed \cite{analytic} that can possibly used as toy-model
backgrounds for computing correlators.
One might hope that the structure coefficients computed for
these backgrounds are interesting, or even recognizable.
In particular, it would be interesting to see if the structure coefficients
resemble those of timelike Liouville theory
\cite{ST,Schomerus,HMW,Giribet} in any way.
Such hopes have yet to be justified.

We have analyzed the three-point function
from the point of view of FRW-CFT, and hence focused on
its property in the FRW patch inside the bubble.
It would be interesting to
investigate its behavior in different regions.
Region D of the Penrose diagram
of figure \ref{BubblePD} is an interesting
region to compare the CDL correlators with
correlators computed around a metastable
vacuum that has not yet decayed.
This is because we can find points that are
arbitrarily far away from the nucleated bubble
in this region.
One might expect that the two correlators
should converge to each other as one travels farther
away from the bubble,
but this is not guaranteed.
It would be worthwhile to check if there is a
discrepancy, and if there is, to understand
its implications properly.

\vspace*{0.1in}

\noindent
{\bf Acknowledgements: }
First and foremost I would like to thank Lenny Susskind for his
support, encouragement and patience throughout the process
of writing this paper, Yasuhiro Sekino for
collaboration on the early stages of this work,
and Wati Taylor and Hong Liu for their support and encouragement
during the course of this work.
I would like to acknowledge that the 
result of section \ref{ss:ancons3}---which was the
crucial step in obtaining the results of this paper---was
jointly obtained with Yasuhiro Sekino.
I would also like to thank Koushik Balasubramanian,
Xi Dong, Ben Freivogel,
Alan Guth, Daniel Harlow, Olaf Hohm, Bart Horn,
Hong Liu, Yasuhiro Sekino,
Douglas Stanford, Lenny Susskind, Richard Melrose,
Wati Taylor and Erik Tonni for useful discussions which this paper
would have been impossible without.
I would like to thank the Stanford Institute for Theoretical Physics,
the Perimeter Institute for Theoretical Physics and the organizers of
Holographic Cosmology 2.0 and the organizers of
Fundamental Issues in Cosmology for their
hospitality during the various stages of this work.
This work was supported in part by funds provided by the
U.S. Department of Energy (D.O.E.) under cooperative research
agreement DE-FC02-94ER40818.
I also acknowledge support as a String Vacuum Project Graduate Fellow,
funded through NSF grant PHY/0917807.

%--------------- ARTICLE ---------------------------

\appendix
\section{Proof of the Completeness Relation} \label{ap:complete}

We prove \eq{delta} for radial potentials $U(X)$
whose continuous eigenfunctions satisfy the regularity conditions.
Many of the results on one-dimensional scattering we use in
this section can be found in \cite{FSSY,Barton}.

Recall that by properties of $a(X)$,
\be
U(X) \rightarrow 1 \quad \text{for } X\rightarrow \pm \infty\,.
\ee
and hence there exist a continuum of states
labelled by real number $k$,
\be
(-\p_X^2 + U(X)) \Psi_k (X) = (k^2+1) \Psi_k (X)\,.
\ee
$\Psi_k$ are defined to be the solutions that behave asymptotically as
\begin{align}
\Psi_k &\rightarrow \begin{cases}
e^{ikX} + \sR(k) e^{-ikX}   & (X \rightarrow -\infty) \\
\sT(k) e^{ikX}  & (X \rightarrow \infty)
\end{cases}
\end{align}
$\Phi_k$ are defined to be the solutions that behave asymptotically as
\begin{align}
\Phi_k &\rightarrow \begin{cases}
\sT(k) e^{-ikX}   & (X \rightarrow -\infty) \\
e^{-ikX} - {\sR(-k) \sT(k) \ov \sT(-k)} e^{ikX}   & (X \rightarrow \infty)
\end{cases}
\end{align}

We say $U(X)$ is regular when $U(X)$ and its eigenfunctions
satisfy the following conditions:
\ben
\item The poles of $\Phi_k$, $\Psi_k$
and $\sT(k)$ with respect to $k$
in the upper-half of the complex $k$ plane coincide
and are simple.
\item The number of such poles are finite.
\item All these poles $iz$ lie on the
imaginary axis and correspond to unique bound states
of energy $(1-z^2)$.
\item $\Phi_k / \sT(k)$ does not have a pole in the
upper-half of the complex $k$ plane.
\item $U(X)$ approaches $1$ as $X \ra -\infty$ ``rapidly."
\een

From standard scattering theory, we know that $\Psi_k$, $\Phi_k$ for $k>0$
together with the bound states $u_{iz}$ form a complete orthonormal
basis of functions on the real line.
Therefore the delta function can be written as
\be
\delta(X-X') = \int_{0}^\infty {dk \ov 2\pi}
\Phi_{k} (X) \Phi_{k}(X')^* +
 \int_{0}^\infty {dk \ov 2\pi} \Psi_{k} (X) \Psi_{k}(X')^*+
 \sum_{iz} u_{iz}(X) u_{iz}(X') \label{delinit}
\ee
where we sum over $iz$ which are poles of $\sT$ in the upper-half plane.
Using the relations \eq{psiphirel1} and \eq{psiphirel2} we obtain
\begin{align}
\begin{split}
  &\int_0^{\infty} {dk \ov 2\pi} \Psi_k (X) \Psi_k (X')^*+  \int_0^{\infty} {dk \ov 2\pi} \Phi_k(X) \Phi_k (X')^*  \\
= &\int_0^{\infty} {dk \ov 2\pi} \Psi_{k} (X) \Psi_{-k} (X')+  \int_0^{\infty} {dk \ov 2\pi} \Phi_{k} (X) \Phi_{-k} (X')\\
= &\int_0^{\infty} {dk \ov 2\pi} \Psi_{k} (X)  ({1\ov \sT(k)} \Phi_{k}(X') + {\sR(-k) \ov \sT(-k)} \Phi_{-k} (X'))\\
 &+\int_0^{\infty} {dk \ov 2\pi} \Phi_{-k} (X') ({1 \ov \sT(-k)} \Psi_{-k}(X) - {\sR(-k) \ov \sT(-k)} \Psi_{k}(X) )\\
= &\int_{-\infty}^{\infty} {dk \ov 2\pi}  {\Phi_k (X) \ov \sT(k)}\Psi_k (X') \,.
\end{split}
\end{align}
Therefore \eq{delinit} implies that
\be
\delta(X-X') =\int_{-\infty}^{\infty} {dk \ov 2\pi}  {\Phi_k (X) \ov \sT(k)}\Psi_k (X') +  \sum_{iz} u_{iz}(X) u_{iz}(X')\,.
\label{comp}
\ee

By assumption
\be
u_{iz} (X) \propto \Res{k=iz} \Psi_k (X) \propto \Res{k=iz} \Phi_k (X) \,.
\ee
Also, since the only simple poles of $\Phi_k (X)$ and $\Psi_k (X')$
in the upper half plane are at the poles of $\sT$ on the imaginary axis,
we may write
\be
 u_{iz} (X) u_{iz} (X') = C_{z} \Res{p=iz} \left( {\Phi_k (X) \ov \sT(k)}\Psi_k (X') \right)
\ee
for a finite number of constants $C_z$.
Therefore (\ref{comp}) becomes
\be
\delta(X-X')=\int_{-\infty}^{\infty} {dk \ov 2\pi}  {\Phi_k (X) \ov \sT(k)}\Psi_k (X')
+ \sum_{iz} C_z \Res{k=iz} \left( {\Phi_z (X) \ov \sT(z)}\Psi_z (X') \right) \,.
\label{cj}
\ee

Let us denote the number of poles of $\sT$ in the upper half plane $n$,
and number the poles $iz_1, \cdots,  iz_n$.
Choosing $n$ points $X_m$ and a point $X'$,
$C_{z_q}$ can be obtained by solving the $n$
independent linear equations
\be
 \sum_{q=1}^n  \left[ \Res{k=iz_q} \left( {\Phi_k (X_m) \ov \sT(k)} \Psi_k (X') \right) \right] C_{z_q}
 = - \int_{-\infty}^{\infty} {dk \ov 2\pi} {\Phi_k (X_m) \ov \sT(k)}\Psi_k (X')
\label{eqts}
\ee
for each point $X_m$.
Conversely, if some set of $C_{z_q}$ satisfies the equation (\ref{eqts}) for
at least $n$ points $X_m$ and a point $X'$, they would satisfy equation (\ref{cj}).

We claim the $C_z =-i$ for all the poles.
To show this, we first acknowledge that
\be
\int_C {dk \ov 2\pi}  {\Phi_k (X) \ov \sT(k)} \Psi_k (X')=
 \int_{-\infty}^{\infty} {dk \ov 2\pi}  {\Phi_k (X) \ov \sT(k)}\Psi_k (X') - \sum_{iz} i \Res{k=iz} \left( {\Phi_k (X) \ov \sT(k)}\Psi_k (X') \right)
\label{first}
\ee
where $C$ is the contour that goes along the real axis with a jump over
the upmost pole of $\sT$. This is depicted in figure \ref{f:contourp}.

Now let us examine $\Phi_k / \sT(k)$ at $X \rightarrow -\infty$
and $|k| \rightarrow \infty$.
It is possible to write the asymptotic expansion of $\Phi_k / \sT(k)$ as
\be
{\Phi_k \over \sT(k)} = e^{-ikX} \left(
1 + \sum_{n=1}^\infty {c_n(X) \ov k^n}
\right) \label{asymp}
\ee
where $c_n(X)$ is a function independent of $k$.
Plugging this ansatz in the Schr\"odinger equation, one actually finds that
\begin{align}
c_1(X) &=- {1 \ov 2i} \int_{-\infty}^X dx (U(x)-1) \\
c_{n+1} (X) &= -{1 \ov 2i} \left[ \int_{-\infty}^X dx (U(x)-1) c_n(x) - c_n'(X) \right] \quad n \geq 1 \,.
\end{align}

We say that $(U(X)-1)$ approaches $0$ ``rapidly" as $X \rightarrow -\infty$,
if there exists an $X_B<0$ such that for
all $X <X_B$, $\int_{-\infty}^X dx (U(X)-1)$ and $U'(X)$ are small enough.
By small enough, we mean that for $X <X_B$ the r.h.s. of \eq{asymp}
converges for large $k$. This means that the ansatz is valid, and the identity
\eq{asymp} is well defined for large $k$.
It is easy to check that
\be
U(X) = 1 +Ae^{NX}
\ee
for positive $N$ is rapid enough by explicit evaluation
of coefficients $c_n$.

Then for any $X<X' < X_B <0$,
\be
{\Phi_k (X) \ov \sT(k)} \Psi_k (X') =
e^{i(X'-X)k} \left( 1+ \OO({1 \ov k})\right)
+\sR(k) e^{-i(X+X')k} \left( 1+ \OO({1 \ov k})\right)
\ee
as $|k| \rightarrow \infty$.
Let us integrate this along the contour $C$.
Since $(X'-X)>0$ and $-(X+X')>0$
the contour integral of both terms along the infinite half circle $C'$ in the
upper half plane is zero, {\it i.e.,}
\be
\int_{C'} {dk \ov 2\pi} {\Phi_k (X) \ov \sT(k)} \Psi_k (X') =0 \,.
\ee
By assumption, there are no poles of the integrand inside $C-C'$,
we actually find that
\be
\int_C {dk \ov 2\pi} {\Phi_k (X) \ov \sT(k)} \Psi_k (X') =
\int_{C'} {dk \ov 2\pi} {\Phi_k (X) \ov \sT(k)} \Psi_k (X') =0
\,.
\ee
Hence for any $X < X'<X_B <0$
\begin{align}
\begin{split}
 \int_{-\infty}^{\infty} {dk \ov 2\pi}  {\Phi_k (X) \ov \sT(k)}\Psi_k (X') - \sum_{iz} i \Res{k=iz} \left( {\Phi_k (X) \ov \sT(k)}\Psi_k (X') \right) =0 \,.
\end{split}
\end{align}
Hence by our previous argument, $C_{z_q} =-i$ for all poles in the upper half plane.
Therefore
\begin{align}
\begin{split}
 &\int_C {dk \ov 2\pi}  {\Phi_k (X) \ov \sT(k)} \Psi_k (X') \\
=&\int_{-\infty}^{\infty} {dk \ov 2\pi}  {\Phi_k (X) \ov \sT(k)}\Psi_k (X') - \sum_{iz} i \Res{k=iz} \left( {\Phi_k (X) \ov \sT(k)}\Psi_k (X') \right) \\
=&\delta(X-X')
\end{split}
\end{align}
and our proof is complete.

\section{Properties of $B(\sof{p_i,l_i,m_i})$} \label{ap:b}

Now
\begin{align}
\begin{split}
&B(\sof{p_i,l_i,m_i}) \\
&= \int_0^\infty d R \sinh^2 R \left( \prod_i {N_{l_i} (p_i) q_{p_i l_i} (R) } \right)
 \int d\om \left( \prod_i {Y_{l_i,-m_i} (\om) } \right) \\
&= \int d R \sinh^2 R \left( \prod_i {N_{l_i} (p_i) q_{p_i l_i} (R) } \right) \\
&\quad \times \left[
\left({(2l_1+1)(2l_2+1)(2l_3+1) \ov 4 \pi }\right)^{1/2}
 W^{l_1 l_2 l_3}_{000} W^{l_1 l_2 l_3}_{(-m_1)(-m_2)m_3} 
\right]
\end{split}
\end{align}
where $C^{l_1 l_2 l}_{m_1 m_2 m}$ are Wigner coefficients for $S^2$.

The radial integral is given by
\begin{align}
\begin{split}
&\int d R \sinh^2 R \left( \prod_i {N_{l_i} (p_i) q_{p_i l_i} (R) } \right) \\
&= {1 \ov \pi^{3/2}  (-2i)^{l_1+l_2+l_3} \prod_i \Ga (l_i+3/2) } \\
&\quad \times p_1 p_2 p_3
\int dR \sinh^2 R \prod_i (\sinh R)^{l_i} ({d \ov d \cosh R})^{l_i} {\sin p_i R \ov \sinh R}
\,.
\end{split}
\end{align}
Hence the analytic properties of $B$ with respect to $\sof{p_i}$
are governed by the behavior of
\begin{align}
\begin{split}
p_1 p_2 p_3
\int dR \sinh^2 R \prod_i (\sinh R)^{l_i} ({d \ov d \cosh R})^{l_i} {\sin p_i R \ov \sinh R}
\,.
\end{split}
\end{align}

We first note that near $\sof{p_i}=(0,p_2,p_3)$
\be
B(\sof{p_i,l_i,m_i}) \sim p_1^2 (C + \text{(higher order terms)})\,,
\ee
when $p_2$ and $p_3$ are generic.
In particular, this is always true when $p_2$ and $p_3$ are real.
This is because for $p$ near zero
\begin{align}
\begin{split}
q_{pl} (R) = {2(-2i)^{l} \Ga(l+3/2) \ov \sqrt{\pi} \prod_{j=1}^l j^2} (\sinh R)^l
({d \ov d \cosh R})^l {R \ov \sinh R} + \OO(p) \,
\end{split}
\end{align}
and
\be
N_{l} (p) = {(-1)^l  \Ga(ip+l+1) \Ga(-ip+l+1) \ov 2^{2l+1} \Ga(l + {3 \ov 2})^2  \Ga(ip) \Ga(-ip)}
= {(-1)^l \prod_{j=0}^l j^2 \ov 2^{2l+1} \Ga(l+3/2)^2} p^2 + \OO(p^4) \,.
\ee
This behavior obviously also holds near $p_2 \sim 0$ and $p_3 \sim 0$ also.
In particular, near $\sof{p_i}=(0,0,0)$,
\be
B(\sof{p_i,l_i,m_i}) \sim p_1^2 p_2^2 p_3^2 (C + \text{(higher order terms)})\,.
\ee

We note that $B$ only has codimension-one poles. Since we have shown that
$\sof{p_i}=(0,0,0)$ is a regular point of $B$, all we have to show is that
\begin{align}
\begin{split}
&\int dR \sinh^2 R \prod_i (\sinh R)^{l_i} ({d \ov d \cosh R})^{l_i} {\sin p_i R \ov \sinh R} \\
&= \int dR \sinh^2 R \prod_i (\sinh R)^{l_i} ({1 \ov \sinh R}{d \ov d R})^{l_i} {\sin p_i R \ov \sinh R}
\end{split}
\end{align}
could have at most a codimension-one pole.
Since
\be
{1 \ov \sinh R} = 2(e^{-R}+e^{-2R} + e^{-3R}+\cdots)
\ee
for $R>0$, the integrand can be written as a sum of terms
\be
\sum_\text{signs} \sum_{n_i=0}^\infty c_{\sof{n_i}}(\sof{\pm p_i, l_i}) e^{(-1-\sum_i n_i +i\sum_i \pm p_i )R}
\ee
where $c_{\sof{n_i}}(\sof{\pm p_i, l_i}) $ are polynomials with respect to $\sof{p_i}$.
Therefore the integral is given by
\be
\sum_\text{signs} \sum_{n_i=0}^\infty {c_{\sof{n_i}}(\sof{\pm p_i, l_i}) \ov -i\sum_i \pm p_i +1+\sum_i n_i }
\ee
which clearly can have only codimension-one poles possibly when
\be
i\sum_i \pm p_i
\ee
is a positive integer.

\section{Regularity of Radial Potentials for Massless and Massive Scalars\\
in a Thin-wall CDL Instanton Background}

We show that the radial potentials for the examples in
sections \ref{s:m0thinwall} and \ref{s:m1thinwall} are 
regular. We state, once more, the regularity conditions 
in the notation defined in section \ref{ss:euc2}:
\ben
\item The poles of $\Phi_k$, $\Psi_k$
and $\sT(k)$ with respect to $k$
in the upper-half of the complex $k$ plane coincide
and are simple.
\item The number of such poles are finite.
\item All such poles $iz$ lie on the
imaginary axis and correspond to unique bound states
of energy $(1-z^2)$.
\item $\Phi_k / \sT(k)$ does not have a pole in the
upper-half of the complex $k$ plane.
\item $U(X)$ approaches $1$ as $X \ra -\infty$ ``rapidly."
\een

\subsection{The Scalar Massless in the Flat FRW Patch} \label{aps:m0}

In this section, we verify the regularity
of the radial potential for the
scalar massless in the flat region of a
thin-wall CDL instanton:
\begin{align}
U(X) &= \begin{cases}
1   & (X < X_0) \\
1-{2-m^2 \ov \cosh^2 X}  & (X > X_0)
\end{cases}
\end{align}

As noted in section \ref{ss:euc2}, when $U(X)$
is constant for all $X<X_B$ for some $X_B$, it is regular.
Since $U(X)$ for the scalar massless in the flat region
satisfies this condition, we know that it is regular.
We, however, check the regularity of $U(X)$ in this section
explicitly.

$\Phi_k$ and $\Psi_k$ are explicitly computed in
section \ref{ss:modes}.
Condition 4 has been checked at the end of
this section and condition 5 is trivial as $U=1$ for $X<X_0$.

It is clear from the definitions of $\Phi_k$ and $\Psi_k$
that the first three conditions will be satisfied if
the poles of $\sR$ and $\sT$ coincide at finite points
on in the upper-half plane.
All such poles are automatically on the imaginary axis.
If not, this means that there exists a normalizable eigenfunction
of the Hamiltonian that has an imaginary eigenvalue,
which cannot be the case since the Hamiltonian is Hermitian.
We claim that when the mass $m$ is small
this is indeed the case.

We show this by studying the
reflection and transmission coefficient.
To do so, let us rewrite the expressions \eq{r1} and \eq{t1}
for $\sR$ and $\sT$:
\begin{align}
 \sR(k) &= e^{2ikX_0}
 \left[ {(t-1){ b_k (t) + \epsilon \{ (t-1)(d_k (t)-b_k (t) ) \}  } \ov
 (ik+1) c_k (t) -\epsilon \{ t c_k(t)+{t(t-1)  \ov (1-\epsilon)}c_k'(t) \}  } \right] \\
 \sT(k) &= {ik \ov (ik+1) c_k (t) -\epsilon \{ t c_k(t)+{t(t-1)  \ov (1-\epsilon)}c_k'(t) \}  } 
\end{align}
where
\begin{align}
b_k (x) &= F(-\nu+1,\nu+1;1-ik;x)  \\
c_k (x) &= F(-\nu,\nu;1-ik;x)\\
c'_k (x) &= {d \ov dy}F(-\nu,\nu;1-ik;y) |_{y=x}\\
d_k (x) &= F(-\nu,\nu+1;1-ik;x)
\end{align}
$t$ is defined as
\be
t \equiv {L \ov 2} \equiv {e^{-X_0} \ov 2 \cosh X_0} \,.
\ee
It is clear that $0<t<1$.

The hypergeometric functions $b_k, c_k, c_k'$ and $d_k$ do not have poles
with respect to $k$ in the upper-half plane---the poles are situated at $k=-i,-2i,\cdots$.
In fact, these functions are bounded in the upper-half $k$ plane since all
of them are analytic and
\be
F(a,b;1-ik,x) \rightarrow 1,\quad\text{for }|k| \rightarrow \infty
\ee
in the upper-half plane.
Therefore when the mass $m$ is small---and hence $\epsilon \sim m^2/3$ is
small---the order $\epsilon$ pieces in the numerator and denominator
of $\sR$ and $\sT$ can only shift the potential zeros or poles
of the numerator or denominator by a very small amount.
Therefore all the poles of $\sR$ and $\sT$ in the upper-half plane
coincide and can be found near the zeros of
\be
(ik+1) c_k (t) = (ik+1)F(-\nu,\nu;1-ik;t)\,.
\ee

We claim that $c_k (t)$ does not have any zeros in the upper-half plane
when the mass $m$ is small.
When $m=0$, $\nu=1$ and hence
\be
|c_k (t)| = |F(-1,1;1-ik;t)| = \left| 1 -{t \ov 1-ik} \right| > 1-t\,.
\ee
in the upper-half of the $k$ plane.
Since $0<t<1$, this is never zero for $k$ in the upper-half plane.
Meanwhile,
\be
\p_\nu c_k (t) |_{\nu=1} = \p_\nu F(-\nu,\nu;1-ik;t)  |_{\nu=1} 
\ee
is also bounded in the upper-half plane.
This can be shown by differentiating the defining
equation for the hypergeometric function
\begin{align}
\begin{split}
F(-\nu,\nu;1-ik;t)= \sum_{n=0}^\infty  {(-\nu)_n (\nu)_n \ov n! (1-ik)_n} t^n \,,
\end{split}
\end{align}
where $(x)_n$ is defined to be
\be
(x)_n \equiv x(x+1) \cdots (x+n-1) \,.
\ee
One finds that
\begin{align}
\begin{split}
\p_\nu F(-\nu,\nu;1-ik;t)  |_{\nu=1}
&= -{2 \ov 1-ik} t + \sum_{n=1}^\infty {n! \ov (1-ik)(2-ik) \cdots (n+1-ik)} t^{n+1}\\
\Rightarrow
\left| \p_\nu F(-\nu,\nu;1-ik;t)  |_{\nu=1} \right|
&\leq 2t + \sum_{n=1}^\infty \left| {n! \ov (1-ik)(2-ik) \cdots (n+1-ik)} \right| t^{n+1}\\
&\leq 2t + \sum_{n=1}^\infty { t^{n+1} \ov n+1 } = \ln(1-t)+t \,.
\end{split}
\end{align}
in the upper-half plane.

Since $|c_k (t)| \geq (1-t)$ when $\nu=1$ and $|\p_\nu c_k (t)|$ is bounded
in the upper-half plane at this value of $\nu$, when the mass $m^2$
is sufficiently small as $\nu \sim 1-m^2/3$, $c_k (t)$ is never zero in
the upper-half plane.
Therefore the only zero of $(ik+1) c_k (t)$ in the upper-half plane
is $k=i$.

Putting everything together, we conclude that the only pole of
$\sR$ and $\sT$ in the upper-half plane coincide and is
near $k=-i$. We thereby conclude the verification
of the regularity of $U(X)$.
We note that the pole in the upper-half plane
should be slightly below $(-i)$, as a small mass gives a positive
contribution to the potential, shifting the bound state energy
in the positive direction.

\subsection{The Massive Scalar} \label{aps:m1}

In this section, we verify the regularity
of the radial potential for the
massive scalar in a thin-wall CDL instanton:
\begin{align}
U(X) &= \begin{cases}
1+\mu^2 e^{2X}   & (X < X_0) \\
1-{2-m^2 \ov \cosh^2 X}  & (X > X_0)
\end{cases}
\end{align}

$\Phi_k$ and $\Psi_k$, as well as the reflection and
transmission coefficients are explicitly computed in
section \ref{ss:m1modes}.
Condition 4 has also been checked at the end of this section.
Condition 5 is also true for $U(X)$ as potentials
with exponential decay are rapid enough.
This fact has been commented on in appendix \ref{ap:complete}.

The fact that the poles of $\Phi_k$, $\Psi_k$ and $\sT(k)$
coincide in the upper-half plane
can be shown by using the following trivial fact.
$\Phi_k(X)$($\Psi_k(X)$) has a pole for all $X \in (a,b)$
if and only if $\Phi_k(X)$($\Psi_k(X)$) has a pole for all $X$.\footnote{This
fact can be checked explicitly in our case by using properties
of hypergeometric and Bessel functions.}
Hence we can look on either side of the wall $X=X_0$
to track the analytic behavior of the eigenfunctions.
\begin{align}
{\Phi_k(X) \ov \sT(k)} &=  \psi_{-k} (X) = ({\mu \ov 2})^{ik} \Ga(-ik+1) I_{-ik} (\mu e^{X})
&\text{for }X<X_0\\
{\Psi_k(X) \ov \sT(k)} &= \phi_k (X) = e^{ikX} F(-\nu,\nu+1;1-ik;{1-\tanh X \ov 2})
&\text{for }X>X_0
\end{align}
It is clear that these functions do not have any poles in the upper-half plane.
Hence the only possibility for a mismatch in poles of $\Phi_k$, $\Psi_k$ and $\sT(k)$
is when $\sT(k)$ has a pole and either $\psi_{-k}$ or $\phi_k$ has a zero or
vice versa.
For small mass $m$ we can show this cannot happen
by methods employed in section \ref{aps:m0}.
Namely, we can show that $\psi_{-k}$($\phi_k$) does not have any zeros(poles)
in the upper-half plane when $m=0$. Then we can show that $\psi_{-k}$($\phi_k$)
changes smoothly with respect to
$m$ around $m=0$ when $k$ is in the upper-half plane.
Thereby we can prove that condition 1 holds
for small mass.

We can show there is a unique pole of $\sT(k)$ in the upper-half plane
near $k=i$ when $m$ is small
by methods used in \ref{aps:m0},
thereby confirming condition 2.
We do not reproduce the proof as it is a mere repetition
of previous arguments.
We note the interesting fact that $\sR(k)$ actually has an infinite number
of poles in the upper-half plane.
Recall that
\be
\Psi_k (X)= \psi_k (X) + \sR(k) \psi_{-k}(X)   \qquad (X < X_0)\,.
\ee
Now $\psi_{-k}$ does not have any poles in the upper-half plane,
but $\psi_{k}$ does. The poles of $\sR(k)$ precisely cancel those poles.

One can show by contradiction that the unique pole should be
on the imaginary axis.
Assume that the pole of $\sT(k)$ is not pure imaginary.
By examining the residue of $\Psi_k$,
we find a normalizable eigenfunction
of the Hamiltonian that has an imaginary eigenvalue.
This cannot be the case since the Hamiltonian is Hermitian.
If the pole is on the imaginary axis, the residue of $\Psi_k$
yields the corresponding bound state wavefunction.
Condition 3 is verified.

%--------------- Bibliography ---------------------------

\end{document}